\documentclass[aps,prb,twocolumn,citeautoscript,footinbib,superscriptaddress]{revtex4-1}

\usepackage{graphicx}
\usepackage{enumitem}
\usepackage{gensymb}
\usepackage{amsmath}
\usepackage{amssymb}
\usepackage{dcolumn}
\usepackage{color}
\begin{document}

\title{Evidence for Stable Square Ice from Quantum Monte Carlo}

\author{Ji Chen}
\thanks{These two authors contributed equally}
\affiliation{Department of Physics and Astronomy, University College London, Gower Street, London WC1E 6BT, U.K.}
\affiliation{London Centre for Nanotechnology, 17-19 Gordon Street, London WC1H 0AH, U.K.}
\affiliation{Thomas Young Centre, University College London, 20 Gordon Street, London, WC1H 0AJ, U.K.}

\author{Andrea Zen}
\thanks{These two authors contributed equally}
\affiliation{Department of Physics and Astronomy, University College London, Gower Street, London WC1E 6BT, U.K.}
\affiliation{London Centre for Nanotechnology, 17-19 Gordon Street, London WC1H 0AH, U.K.}
\affiliation{Thomas Young Centre, University College London, 20 Gordon Street, London, WC1H 0AJ, U.K.}

\author{Jan Gerit Brandenburg}
\affiliation{London Centre for Nanotechnology, 17-19 Gordon Street, London WC1H 0AH, U.K.}
\affiliation{Thomas Young Centre, University College London, 20 Gordon Street, London, WC1H 0AJ, U.K.}
\affiliation{Department of Chemistry, University College London, 20 Gordon Street, London WC1H 0AH, U.K.}

\author{Dario Alf\`e}
\affiliation{Department of Physics and Astronomy, University College London, Gower Street, London WC1E 6BT, U.K.}
\affiliation{London Centre for Nanotechnology, 17-19 Gordon Street, London WC1H 0AH, U.K.}
\affiliation{Thomas Young Centre, University College London, 20 Gordon Street, London, WC1H 0AJ, U.K.}
\affiliation{Department of Earth Sciences, University College London, Gower Street, London WC1E 6BT, U.K.}

\author{Angelos Michaelides}
\email{angelos.michaelides@ucl.ac.uk}
\affiliation{Department of Physics and Astronomy, University College London, Gower Street, London WC1E 6BT, U.K.}
\affiliation{London Centre for Nanotechnology, 17-19 Gordon Street, London WC1H 0AH, U.K.}
\affiliation{Thomas Young Centre, University College London, 20 Gordon Street, London, WC1H 0AJ, U.K.}

\begin{abstract}
Recent  experiments on ice formed by water under nanoconfinement provide evidence for a two-dimensional (2D) `square ice' phase.
However, the interpretation of the experiments has been questioned and the stability of square ice has become a matter of debate.
Partially this is because the simulation approaches employed so far
(force fields and density functional theory) struggle to
accurately describe the very small energy differences between the relevant phases.
Here we report a study of 2D ice using 
an accurate wave-function based electronic structure approach, namely Diffusion Monte Carlo (DMC).
We find that at relatively high pressure square ice is indeed the lowest enthalpy phase examined, supporting the initial experimental claim.
Moreover, at lower pressures a `pentagonal ice' phase (not yet observed experimentally) has the lowest enthalpy, 
and at ambient pressure the `pentagonal ice' phase is degenerate with a `hexagonal ice' phase.
Our DMC results also allow us to evaluate the accuracy of various density functional theory exchange correlation functionals and force field models, and in doing so we extend the understanding of how such methodologies perform to 
challenging 2D structures presenting dangling hydrogen bonds.
\end{abstract}

\maketitle

Recent transmission electron microscopy (TEM) measurements and classical molecular dynamics simulations 
report that a new two dimensional (2D) square phase of ice forms~\cite{algara-siller_square_2015}.
This phase is not part of the bulk ice phase diagram, but
it was suggested that it is stabilized under confinement because of 
lateral pressure 
(estimated to be in the gigapascal, GPa, regime), arising from the van der Waals (vdW) attraction between
the graphene sheets.
However, these experiments have been questioned~\cite{zhou_observation_2015},
with it even being suggested that it is sodium chloride contamination and not ice that is responsible for
the square symmetry observed.
So far, it is not clear under what conditions (if any) square ice is stable.

\begin{figure}[tbhp]
\includegraphics[width=2.9in]{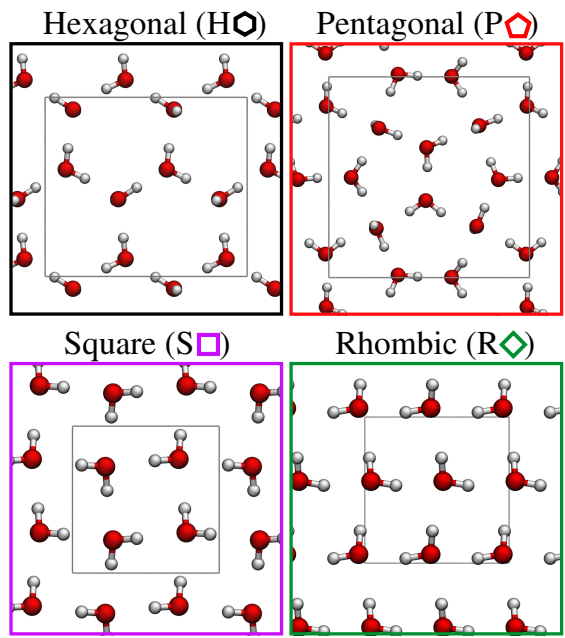}
\caption{
The four most relevant 2D ice structures considered as part of this study:
hexagonal (H),
pentagonal (P),
square (S), and
rhombic (R).
The gray boxes represent the primitive unit cells.
}
\label{fig:str}
\end{figure}

Theoretical investigations of the stability of confined 2D ice at high lateral pressures can,
in principle, help in disentangling this issue and in complementing experimental findings. 
From a theoretical perspective, the prediction of square 2D ice 
can be traced back to 
Nagle's 1970s
`unit model' of ice \cite{NAGLE1979317}.
However, later atomistic force field (FF) simulations found
that 2D ice prefers a buckled rhombic structure \cite{zangi_monolayer_2003, koga_phase_2005}.
More recently, density functional theory (DFT) based investigations~\cite{corsetti_structural_2016, chen_two_2016, corsetti_enhanced_2016,roman_polymorphism_2016} have been performed.
However these have produced qualitatively different results 
depending on the precise details of the calculations and, in particular, on the choice of exchange-correlation (XC) functional.
For instance, \citet{chen_two_2016} found hexagonal and pentagonal structures to be stable phases 
at low pressures (Fig.~\ref{fig:str}).
They also found that square ice is only stable in the GPa pressure regime and that 
even at these pressures it is only favored enthalpicly by $<$ 10 meV/H${_2}$O (1/4 kcal/mol).
Meanwhile, DFT results of 
\citet{corsetti_structural_2016} showed that the square ice structure is more stable than the hexagonal phase at all pressures.
These disparate findings raise serious questions about the reliability of the adopted computational approaches (both DFT and FF) applied so far to 2D ice. 
Indeed this is exemplary of a broader longstanding issue: the water and ice phase diagram is extremely challenging for any computational approach, because there can be competing phases within an energy range of only a few tens of meV/H${_2}$O \cite{Ice:prl2011}.
Achieving this accuracy is often beyond the capabilities of DFT XC functionals and most FF approaches \cite{gillan_perspective:2016,cisneros_modeling_2016}.
Therefore, a study of 2D ice with a more accurate theoretical method is needed. 

The development and application of electronic structure approaches with meV accuracy is now a thriving area of condensed matter research (see e.g. Refs.~\cite{FCIQMC:Nat2013,RPA_Schimka:NatMatt2010,michaelides_jcp_preface, chan_benzene}). 
Of the various methods available diffusion Monte Carlo (DMC) is particularly attractive \cite{foulkes01,Wagner:prb2014,Morales:bulkwat:2014, Benali:2014, Mazzola:prl2015,Zen:liqwat2015,noncov:chemrev2016}.
First, DMC has already been shown to offer the requisite accuracy for bulk ice phases by producing
results in excellent agreement with experiment ~\cite{Ice:prl2011,santra_on_2013,Quigley:Ice_0_i_Ih:jcp2014,gillan_perspective:2016}. 
Second, thanks to recent improvements in computational efficiency \cite{sizeconsDMC}, 
it is now possible to obtain converged results
on the large unit cells that must be considered when treating 2D ice with a many-body 
electronic structure approach.
With this in mind, herein we report a DMC
study of 2D ice. 
Our DMC calculations reveal that hexagonal and pentagonal phases are indeed the
most stable 2D ice phases identified at low pressures.
Perhaps of more interest though, DMC calculations at 2 GPa clearly support the existence of 
the square ice at high lateral pressure.
As a further step we then use our DMC reference data to understand how the much more widely
used DFT and FF approaches perform for such systems, so as to provide guidance for future
studies on 2D ice and
interfacial water. 
From this, several DFT XC functionals which perform well are identified and we also find that
commonly used FFs such as SPC/E~\cite{berendsen_missing_1987} 
and TIP4P~\cite{jorgensen_comparison_1983} tend to overstabilize high density phases;
helping to explain why such
phases have been widely observed in FF studies.

Previous DFT studies suggest that four structures -- hexagonal, pentagonal, square, and rhombic (see Fig.~\ref{fig:str}) -- are the most stable monolayer ice structures \cite{chen_two_2016}, so these are the focus of the current study.
In the rhombic and square structures the water molecules are fourfold coordinated, as in bulk ice, but are arranged in the plane. 
The pentagonal and hexagonal structures have water molecules arranged similarly to a cut in bulk ice, so that there are some water molecules that are not fourfold coordinated and some with dangling hydrogen bonds.
Specifically, in the hexagonal structure all water molecules are threefold coordinated and 1/2 of them have one dangling hydrogen bond; 
in the pentagonal structure 1/3 of the water molecules have one  dangling hydrogen bond and are fourfold coordinated, the remaining molecules are threefold coordinated and have no dangling bonds.
The {\sc casino} code~\cite{casino} has been used for the DMC calculations
using 
Dirac-Fock pseudopotentials~\cite{trail05_NCHF, trail05_SRHF} with the locality approximation~\cite{mitas91}.
Slater-Jastrow trial wave functions with single Slater determinants were used and the single particle orbitals obtained from DFT-LDA plane wave calculations reexpanded in terms of B-splines~\cite{alfe04}.
We used the DMC algorithm with the prescriptions of \citet{sizeconsDMC}, which allows a DMC time step as large as 0.02~a.u. to be used with negligible time step errors. 
DMC calculations have been performed on structures optimized with DFT at a range of lateral pressures and under a uniform 2D confining potential fitted to DMC for values for the water-graphene interaction ~\cite{ma_adsorption_2011}.
See Ref. \onlinecite{chen_two_2016} and the supplemental material therein for details of the confining potential used and discussions on the impact of using explicit graphene as the confining material.
Structures used for DMC calculations have been obtained by performing geometry optimizations with the optPBE-vdW functional~\cite{klimes_chemical_2010}.
Additional calculations have also been performed on selected structures obtained from the revPBE-vdW functional~\cite{klimes_chemical_2010}, leading to similar results, as detailed in the Supplemental Material (S.M.)~\cite{SI}. 
Further details of the set ups used in the DMC, DFT and FF calculations are also included in the S.M.

\begin{figure}[htb]
\includegraphics[width=3.in]{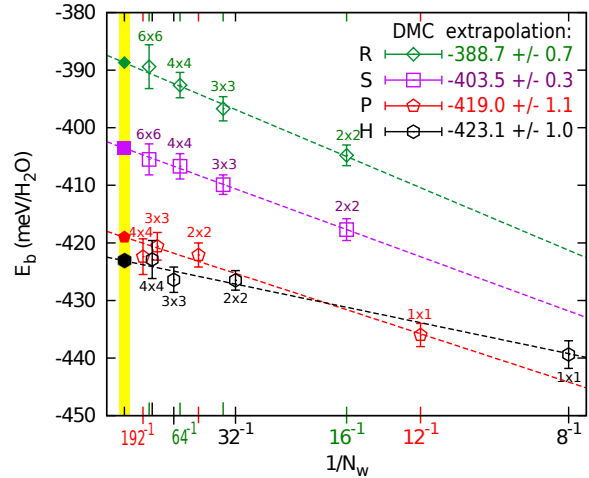}
\caption{
DMC binding energies, $E_\text{b}$, for the 0 pressure structures of free standing 2D ice in the
hexagonal (H), pentagonal (P), square (S), and rhombic (R) structures (see Fig.~\ref{fig:str}) as a function of the inverse number of water molecules in the simulation cell, $1/N_w$. 
Empty symbols correspond to DMC evaluations of $E_\text{b}$, error bars are one standard deviation.
For each point we indicate the supercell size in terms of the primitive unit cell. 
The filled symbols on the yellow region are the extrapolated values, $E_\infty$, from a fit with the function
$E(N_w) = E_\infty - c/N_w$; values and errors of the fit are reported on the top right corner. 
}
\label{fig:qmc}
\end{figure}

In order to assess the stability of the different 2D ice structures, 
it is necessary to consider their enthalpy ($H$) at 0 K~\cite{SI}.
The enthalpy is the sum of the binding energy, $E_\text{b}$  \footnote{ $E_\text{b}$ is calculated as the difference between the energy per water in the free standing 2D ice sheet minus the energy of an isolated water molecule, see S.M.}, the confinement energy, $E_{\rm conf}$,  the pressure volume work ($PV$), and the zero point energy (ZPE).
To begin, we first discuss the accurate evaluation of $E_\text{b}$ using DMC.
In DMC, and other many-body methods, finite size (FS) errors can be sizeable unless large supercells and/or correction terms are considered~\cite{Lin:qmctwistavg:pre2001,Chiesa:size_effects:prl2006,KZK:prl2008}.
Since we are concerned with very small energy differences between the various phases,
we have carefully addressed this issue with a series of calculations for increasing supercell size.
Supercells with 8 up to as many as 192 water molecules were considered and for each system considered DMC simulations were run until the stochastic error of $E_\text{b}$ was $\le$~3~meV/H$_2$O.
The results obtained are plotted in Fig.~\ref{fig:qmc} as a function of the inverse number of water molecules in the simulated supercell, $1/N_w$.
From these calculations we have extrapolated $E_\text{b}$ for the different structures to infinite system size.
The extrapolated binding energies are summarized in Table~\ref{tab:qmc} 
from where it can be seen that
the hexagonal phase has the largest binding energy and the pentagonal phase
is only  marginally (4 meV/$\text{H}_2\text{O}$) less stable.
These calculations also reveal that the FS error leads to an overestimate of $E_\text{b}$ by more than
10~meV/$\text{H}_2\text{O}$ in the smallest cells ($N_w<20$),
and the relative energies between the different structures are fairly insensetive to the size of cell used.
We note that the $N_w\to \infty$ extrapolations are obtained by assuming that FS errors are proportional to $1/N_w$.
In the S.M. we show that other choices do not alter the extrapolated values of $E_\text{b}$~\cite{SI}. 

\begin{table}[htdp]
\caption{
Properties of the various 2D ice structures at 0 pressure and 2 GPa.
The values reported are the binding energy $E_\text{b}$ (meV/H$_2$O) obtained from DMC,
the confinement energy $E_{\rm conf}$ (meV/H$_2$O),
the lateral area per water $A$ (\AA{}$^2$/H$_2$O), 
the pressure volume work $PV=P\times A\times w$ (meV/H$_2$O; $w$ is the width of the confinement),
the zero point energy ($ZPE$) (meV/H$_2$O),
and the enthalpy $H$ (inclusive of $ZPE$).
Also reported is the enthalpy difference with respect to the most stable structure:
$\Delta_\text{pen} = (H) - (H)[\text{pentagonal}]$ at 0 pressure,
$\Delta_\text{sq} = (H) - (H)[\text{square}]$ at 2 GPa.
All the DMC stochastic errors associated with $E_\text{b}$ are $\le 3$~meV/H$_2$O.
$E_{\rm conf}$, $A$, $PV$, and $ZPE$ were obtained using DFT with the optPBE-vdW functional.
See S.M. for additional details.
}
\begin{center}
\begin{tabular}{  | l |  c c c c c c c | c c c c c c c | }
\hline
 0 pressure &	$E_\text{b}$& 	$E_{\rm conf}$ &$A$	& $PV$ 	& $ZPE$	& $\textbf{H}$	&	$\Delta_\text{pen}$	\\
\hline																					
Hexagonal	&	-423 & 21 &	9.728 	&	0	& 676	& \textbf{274}	&	\textbf{1}	\\
Pentagonal	&	-419 & 20 &	8.635 	&	0	& 672	& \textbf{273}	&	\textbf{0}	\\
Square		&	-404 & 18 & 	7.974	&	0	& 672	& \textbf{286}	&	\textbf{13}	\\
Rhombic		&	-389 & 37 & 	7.999	&	0	& 667	& \textbf{315}	&	\textbf{42}	\\
\hline
	2 GPa	&			\multicolumn{6}{c}{  } &	$\Delta_\text{sq}$		\\
\hline																					
Pentagonal	& -380	& 52  & 7.710	&	577	&	674 & \textbf{923}	&	\textbf{28}	\\
Square		& -374	& 47  & 7.193	&	539	&	683 & \textbf{895}	&	\textbf{0}	\\
Rhombic		& -385	& 99  & 6.905	&	517	&	681 & \textbf{912}	&	\textbf{17}	\\
\hline
\end{tabular}
\end{center}

\label{tab:qmc}
\end{table}

With the binding energies obtained, the enthalpies can be further calculated considering $E_{\rm conf}$, $PV$, and ZPE.
Each of these additional terms has been obtained with DFT and upon putting everything together we find that at vanishing pressure the pentagonal and hexagonal structures have essentially the same enthalpy. 
The values obtained come within 1~meV/$\text{H}_2\text{O}$ of each other which is 
within the stochastic errors of our DMC simulations (3 meV/$\text{H}_2\text{O}$). 
Although it is hard to differentiate the pentagonal and the hexagonal structures at vanishing pressure,
based on the enthalpies obtained 
we would expect the higher density pentagonal phase to become 
more stable than the hexagonal phase at small applied pressure, 
as first proposed using DFT \cite{chen_two_2016}. 
The square and rhombic structures are 13 and 42 meV/$\text{H}_2\text{O}$ higher in enthalpy, respectively, than the pentagonal structure at zero pressure.
Note that the rhombic structure is buckled, thus also the confining energy penalizes it compared to the other structures.

In the 2D ice phase diagram, 
the question of the stability of square ice is 
particularly interesting because it is arguably the only experimentally observed 2D ice
and disagreements exist between FFs and DFT, and also within DFT itself \cite{algara-siller_square_2015,zhou_observation_2015, corsetti_structural_2016,chen_two_2016}.
We have already seen that the square phase is less stable than the pentagonal and hexagonal phases at low pressures.
However, it has been estimated that the pressure in graphene nanocapillaries
can be as large as several GPa due to the van der Waals forces pulling the graphene sheets together \cite{algara-siller_square_2015}.
Therefore, we carried out DMC calculations on structures at 2 GPa.
In Table~\ref{tab:qmc} we report results 
for pentagonal, square and rhombic structures;
the hexagonal structure becomes unstable at 2 GPa.
We find that the square structure is indeed the most stable. 
Its enthalpy is lower than the pentagonal structure by 28 meV/$\text{H}_2\text{O}$
and lower than the rhombic structure by 17 meV/$\text{H}_2\text{O}$.
Although this study has focussed on relative enthalpies at 0 K, we have also estimated the relative free energies of the various phases by taking the vibrational contributions to the free energies into account.
As shown in the S.M. (Table S.II) at 300 K the square phase remains the most stable phase at 2 GPa.

\begin{figure}[h!]
\includegraphics[width=3.1in]{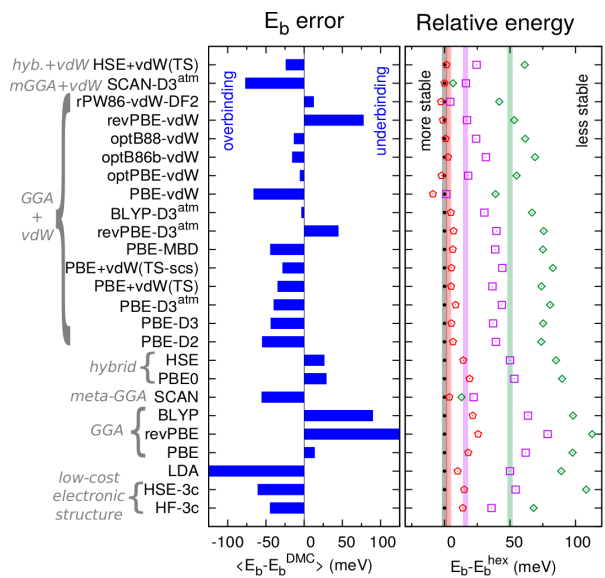}
\caption{ 
Left panel reports the average error (blue bars) 
for the $E_b$ evaluations, taking DMC as the reference, using a selection of XC functionals.
See the S.M. for information on each method.
Right panel reports the energy difference $E_\text{b}-E_\text{b}^\text{hex}$ at 0 pressure for each of the considered methods.
The benchmark DMC evaluations are shown as color-coded bands, the widths of which 
represent the stochastic error (i.e., $\pm \sigma$). Color and symbol conventions are the same as in the previous figures.
Results are on optPBE-vdW optimized structures at 0 pressure.
Similar plots for relaxed structures and for 2~GPa are reported in the S.M.
}
\label{figure3}
\end{figure}

\begin{figure}[h!]
\includegraphics[width=3.in]{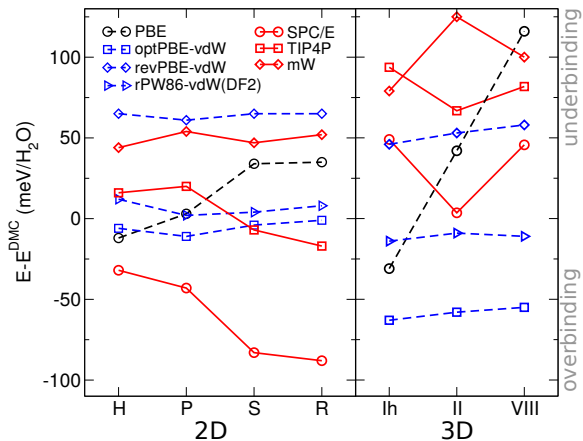}
\caption{
Binding energy difference, $E - E^{\rm DMC}$, with respect to DMC for various DFT XC functionals and FF models for the 2D and some bulk (3D) ice structures at 0 pressure. 
For clarity, only a subset of the FF potentials and DFT XC functionals are plotted here; data
with other approaches is included in the S.M.
Bulk ice results are
taken from Refs. \onlinecite{Ice:prl2011,santra_on_2013} (DMC and DFT) 
and Ref. \onlinecite{sanz_phase_2004} (FF).
In both panels the density of the phases increase from left to right.
}
\label{figure4}
\end{figure}

DMC has helped to clarify the relative stabilities of the various 2D ices at ambient and high pressure.
We now use these DMC benchmarks to understand how various DFT XC functionals and 
FF models perform on such structures.
This is important to establish as DFT and FFs are widely used to examine 2D ice, interfacial, and confined water.
Here we discuss the main indications that come out of these comparisons.
In Fig. \ref{figure3} (left panel) we plot the average error in the binding energy of 2D ice structures as obtained  
from a variety of approaches.
From this we find that several XC functionals, namely HSE-vdW(TS), rPW86-vdW-DF2, optB88-vdW, optB86b-vdW, optPBE-vdW, BLYP-D3$^{\mathrm{atm}}$, and PBE perform well, yielding an average error of $<$ 25 meV/$\text{H}_2\text{O}$.
Apart from binding energies averaged over the four structures, it is also crucial to predict correct relative energies.
As shown by the relative energies of the four
2D ice structures (Fig. \ref{figure3} right), we find that only the vdW inclusive approaches mentioned above provide satisfactory predictions.
The long range part of the vdW force therefore plays 
an important role in 2D ice.
Interestingly, SCAN, a recently developed meta-GGA functional that partially accounts for the medium range vdW force \cite{sun_accurate_2016, brandenburg_benchmark_2016},
does not perform particularly well for these systems and overbinds all phases.

Extending the analysis, in Fig.~\ref{figure4} we plot the energy error of several XC functionals and FFs 
for both 2D and 3D ice.
Results are shown in order of increasing density for both 2D and 3D ice.
On comparing 2D and 3D ice we find that whilst vdW is important in 2D it plays a smaller
role than it is known to play in 3D~\cite{Ice:prl2011,santra_on_2013}.
This can be seen for example by comparing the slopes of the PBE curves in 2D and 3D, 
where it can be seen that the PBE curve in 3D is much steeper.
We also show in the S.M. that the reduced significance of vdW is due to the lower coordination of water molecules in 2D ice than in 3D.
The difference of the vdW contribution in 2D and 3D leads to contrasting performance of some of the 
vdW functionals for 2D and 3D ice,
as shown in Fig.~\ref{figure4}.
Overall we find that the rPW86-vdW-DF2 functional
yields the smallest errors for both 2D and 3D ice structures, and
represents a good choice for future studies of these and related systems.
In Fig. \ref{figure4} we also
present results with several FF models which
have been widely used to study confined water and ice~\cite{koga_first-order_2000, zangi_monolayer_2003, koga_phase_2005, johnston_liquid_2010,algara-siller_square_2015}.
Our calculations show that the SPC/E and TIP4P models overestimate the binding energy of the high density 2D ice structures more than the low density ones.
The difference between the errors on the hexagonal and rhombic structures is about 40 meV/$\text{H}_2\text{O}$ for TIP4P and about 60 meV/$\text{H}_2\text{O}$ for SPC/E.
This error comes from both the Coulomb and the van der Waals components of the potential (Fig. S6).
Results of TIP4P/2005 \cite{abascal_general_2005} and TIP4P/Ice \cite{abascal_a_2005} are also reported in the S.M. and typically they show 
a consistent shift of energy for all structures compared with TIP4P.
Therefore, our results help to explain why rhombic and square structures
are often seen in 2D ice simulations using these models.
We have also performed calculations using the mW model \cite{molinero_water_2009}, a widely used coarse grained model of water.
Although it significantly underestimates the binding energies of the 2D ice structures
by \textit{ca.} 50 meV/$\text{H}_2\text{O}$, it does very well in reproducing the relative energies of 2D ice.
This good performance on the relative energies is also consistent with the fact that a pentagonal
ice structure has been observed in mW simulations of confined water \cite{johnston_liquid_2010}.

To conclude, our large scale DMC simulations of 2D ice reveal that at ambient pressure the most stable structures are hexagonal and pentagonal phases.
In addition, our calculations show that at high pressure 2D square ice is more stable at 0 K than 
the other phases considered.
The data and insight obtained here is important for current understanding of confined 2D ice in general and is also relevant to the TEM measurements of Algara-Siller \textit{et al.}~\cite{algara-siller_square_2015}. 
We note that although our calculations find that square ice is stable they
do not rule out the possibility, as suggested by Zhou \textit{et al.}, that sodium chloride 
could have been observed in the initial measurements~\cite{zhou_observation_2015}.
The influence of finite temperature, growth kinetics, and the finite size of the particles that form have also yet to be investigated in detail.
Clearly more work is needed from experiment and theory in order to
substantiate the existence of square ice, and to potentially observe the predicted pentagonal ice phase.
Finally, from a theoretical point of view, our study reveals the different performance 
of many widely used DFT functionals and FF models on 2D ice.
We find that the role of vdW forces in 2D and 3D ice is different, thus
it is important to consider both 2D and 3D ice in order to reach a consistent picture 
of vdW inclusive XC functionals.
We also rationalize observations in previous FF studies by showing that widely used FF models incorrectly stabilize the high density 2D ice phases. 

\section*{Acknowledgements}
J.C., A.Z. and A.M. are supported by the European Research Council
under the European Union's Seventh Framework Programme
(FP/2007-2013) / ERC Grant Agreement number 616121 (HeteroIce project).
A.Z. and A.M.'s work is also sponsored 
by the Air Force Office of Scientific Research, Air Force Material Command, USAF, under grant number FA8655-12-1-2099.
A.M. is also supported by the Royal Society through a Royal Society Wolfson Research Merit Award.
J.G.B acknowledges support by the Alexander von Humboldt foundation within the Feodor-Lynen program.
We are also grateful for computational resources to ARCHER, UKCP consortium (EP/ F036884/1), 
the London Centre for Nanotechnology, UCL Research Computing, and Oak Ridge Leadership Computing Facility (No. DE-AC05-00OR22725).

\newpage
\setcounter{section}{0}
\renewcommand{\thesection}{S\arabic{section}}%
\setcounter{table}{0}
\renewcommand{\thetable}{S\arabic{table}}%
\setcounter{figure}{0}
\renewcommand{\thefigure}{S\arabic{figure}}%
\section*{Supplemental Material}

\noindent
The material here reported is the following:
computational details in Sec.~\ref{sec:comp}; 
extended data in Sec.~\ref{sec:extdat}; 
additional notes in Sec.~\ref{sec:note};
and structure files used for DMC calculations are described in Sec~\ref{sec:strfiles}.

\section{Additional details of computation}
\label{sec:comp}

\subsection{Evaluated quantities: binding energy, confinement energy and enthalpy}

The {\bf binding energy} $E_b$ (sometimes called cohesive energy) of a given ice structure is defined as
\begin{equation} \label{eq:Eb}
E_\text{b}= E_\textrm{ice} - E_{\text{H}_\text{2}\text{O}} ,
\end{equation}
where
$E_{\text{H}_\text{2}\text{O}}$ is the energy of a water molecule in vacuum
and $E_\textrm{ice}$ is the energy per water in the corresponding 2D ice structure.
The latter is calculated as 
$ \frac{ E_\text{tot,ice} }{ n_{\text{H}_\text{2}\text{O}} }$, where 
$E_\text{tot,ice}$ is the total energy and $n_{\text{H}_\text{2}\text{O}}$
is the number of water molecules in the simulated supercell of 2D ice.

In all our calculations the graphene layers are not simulated with an {\em ab-initio} approach, 
because that would be extremely difficult as the unit cells of graphene and of the 2D ice structures are incommensurable, so huge supercells should be considered (out of the reach for {\em ab-initio} approaches). 
However, in our previous paper we did an extensive set of tests which showed that the {\bf uniform 
confining potential} and explicit graphene confinement led to similar structures~\cite{chen_two_2016}.
Two different types of confinement potentials were used to evaluate the confinement energy and to optimize the 2D ice structures.
The first one is a Morse potential fitted to the quantum Monte Carlo (QMC)
results for the interaction of a water monomer with graphene \cite{ma_adsorption_2011, chen_two_2016}, 
and is defined as
$$
V_\text{Morse}(z) = D((1-e^{-a(z-z_0)})^2-1) ,
$$ 
where $z$ is the distance between the oxygen atom and
the wall, $D = 57.8 $ meV, $a = 0.92~\text{\AA}^{-1}$, $z_0 = 3.85~ \text{\AA}$.
The second one is a Lennard-Jones 9-3 (LJ93) potential, and is defined as
$$
V_\text{LJ93}(z) = \epsilon \left( \frac{5}{12} \left(\frac{\sigma}{z}\right)^9 - \left(\frac{\sigma}{z}\right)^3 \right) ,
$$
where $\epsilon = 21.7 $ meV and $\sigma = 3.0~ \text{\AA}$.
Assume that the 2D ice is in the $xy$-plane and that the two walls are at $z=0$ and $z=w$ ($w$ is the width of the confinement), such that the waters are found in the region $0<z<w$.
At 0 pressure, in order to benchmark force field models, we report results with the LJ93 confining potential with $w=$ 5~\AA, which is also available in force field programs and has been widely used.
As shown in our previous paper~\cite{chen_two_2016}, the 5~\AA~LJ93 confinement leads to very similar results to the Morse potential.
At 2 GPa, as it is relevant to the experiments in graphene confinement, we have used the Morse potential as the confinement to achieve a better description.
At 2 GPa the width of confinement used is 6~\AA. The impact of the width on the stability of 2D ice has also been discussed in our previous paper~\cite{chen_two_2016}.

The corresponding {\bf confinement energy} $E_\text{conf}$ is defined as:
\begin{equation}\label{eq:Econf}
E_\text{conf} = \frac{1}{n_{\text{H}_\text{2}\text{O}}} \sum_{i=1}^{n_{\text{H}_\text{2}\text{O}}} [ V(z_i) + V(w-z_i) ] ,
\end{equation}
where the index $i$ runs over all the water molecules in the simulated supercell, $z_i$ is the $z$ coordinate of the $i$-th water, and the potential $V(\cdot)$ can either be $V_\text{Morse}$  or $V_\text{LJ93}$.
Note that in both cases $E_\text{conf}$ is only a function of the structure, and in particular of the distance between the oxygen atoms and the two walls that realize the confinement.

The {\bf enthalpy} is defined as,
\begin{equation} \label{eq:H}
H = E_\text{b} + E_\text{conf} + P \times A \times w,
\end{equation}
where 
$P$ is the lateral pressure and $A$ is the lateral area per water.

Zero point energy is calculated as,
\begin{equation} \label{eq:zpe}
\text{ZPE}=\sum_{i=1}^{i=N} \frac{1}{2} \hbar \omega_{\Gamma,i},
\end{equation}
and
vibrational free energy at finite temperature is calculated as,
\begin{equation} \label{eq:vib}
G_\text{vib}=\sum_{i=1}^{i=N} \hbar \omega_{\Gamma,i} [\frac{1}{2}+\frac{1}{\text{exp}(\hbar \omega_{\Gamma,i} / k_{B}T)-1}],
\end{equation}
where $\omega_{\Gamma,i}$ is the $i$th gamma point vibrational frequency
calculated using finite displacement method, and $\hbar$ is the reduced Planck constant.

\subsection{DMC}

For all the DMC calculations, we used trial wave-functions of Slater-Jastrow type with single Slater determinants and a Jastrow factor that includes electron-nucleus, electron-electron and electron-electron-nucleus interaction terms.
The single particle orbitals are obtained from DFT-LDA plane-wave calculations, performed with {\sc pwscf}~\cite{pwscf}, using a plane-wave cutoff of 600 Ry and re-expanded in terms of B-splines~\cite{alfe04} with the natural grid spacing $a=\pi/G{\rm max}$,
where $G{\rm max}$ is the modulus of the longest G vector in the PW expansion.
Dirac-Fock pseudopotentials~\cite{trail05_NCHF, trail05_SRHF} and  the locality approximation~\cite{mitas91} are used.
With this wave-function ansatz, the variance of the VMC local energies per molecule is of  $\sim 0.3$~Ha$^2$ in all the considered systems, included the isolated molecule.
2D periodicity is used in all calculations of 2D ice structures, whereas for the single molecule open boundary conditions are used.

\begin{figure}[h!]
\begin{center}
\includegraphics[width=3.3in]{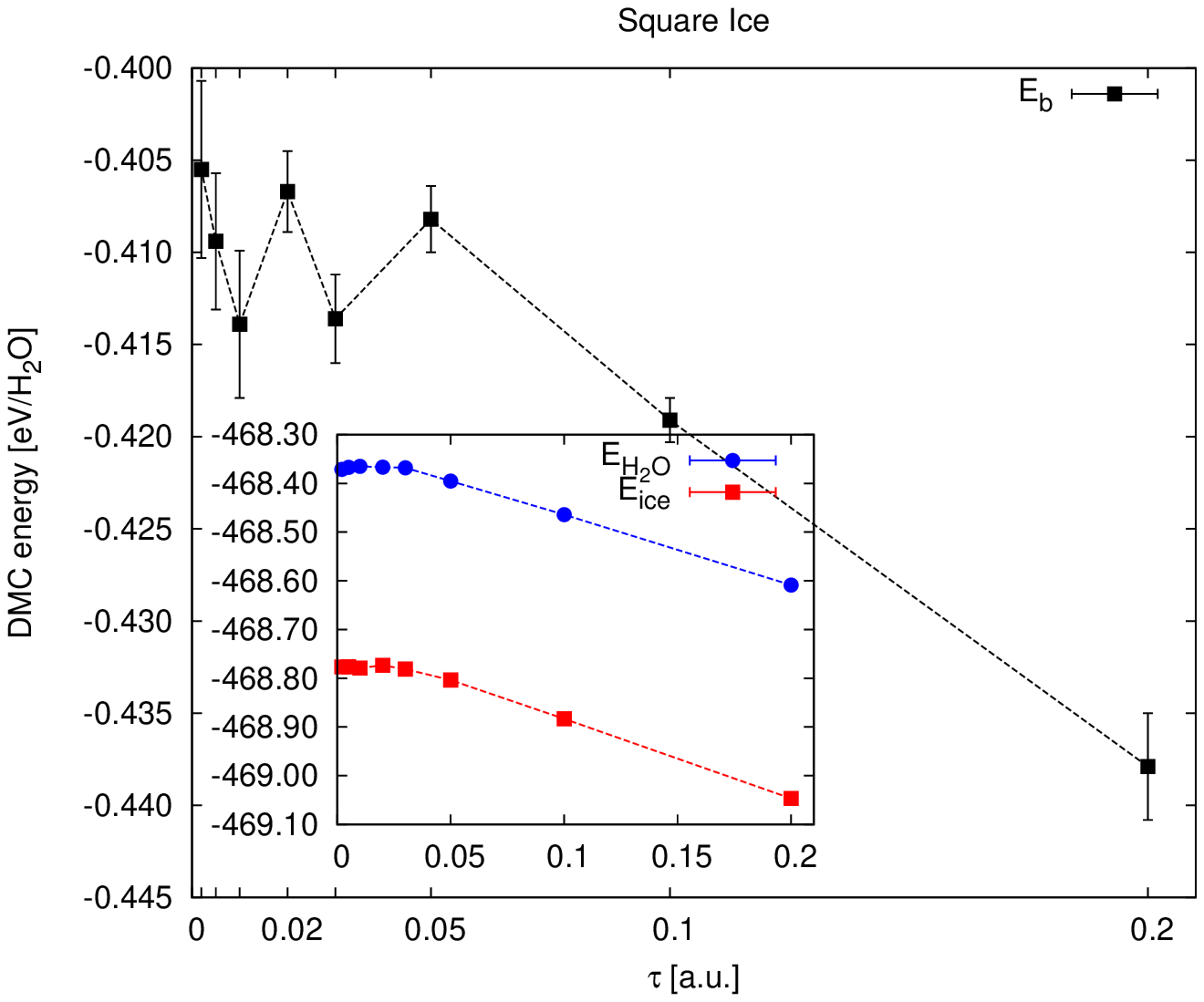}
\end{center}
\caption{
Binding energy $E_b$ of square ice, calculated using DMC for a supercell with 64 waters, as a function of the DMC time-step $\tau$. 
In particular, we performed calculations at $\tau = 0.2, 0.1, 0.05, 0.03, 0.02, 0.01, 0.005, 0.002 $ a.u. 
Error bars correspond to one standard deviation in the energy evaluation.
(Inset) Absolute energy (per water molecule) of an isolated water molecule (blue) and of the square ice (red) as a function of $\tau$. Error bars are smaller than the point size.
}
\label{fig:qmc:tau}
\end{figure}

Based on previous work, we know that one of the most challenging parameters to choose is the DMC time-step $\tau$; 
exact results are obtained for $\tau\to 0$, but the computational cost is $\propto 1/\tau$.
In previous DMC investigations on water systems $E_b$ was evaluated with a $\tau$ as small as 0.002~a.u. in order to get rid of undesired time-step errors (see S.I. of Ref.~\onlinecite{Ice:prl2011}). 
Recently,
\citet{sizeconsDMC} have introduced some improvements to the DMC algorithm that drastically reduces finite time-step errors.
Thus, in this work we used the prescriptions of Ref.~\onlinecite{sizeconsDMC} and a $\tau=0.02$~a.u., which guaranties a negligible time-step error.
In order to show this, we considered the square 2D ice structure (see Fig.~1 in main paper) 
to evaluate the extent of time-step errors.
The unit cell includes four water molecules, and here we consider a $4 \times 4$ supercell, for a total of 64 waters in the system.
At the VMC level of theory the evaluated binding energy is -0.108(4)~eV, that is severely underestimated  (by a factor 4) with respect to the DMC evaluations.
In Fig.~\ref{fig:qmc:tau} we display the energy of the isolated water molecule, as well as the energy per water in the square structure, as a function of the DMC time-step $\tau$. 
No sizeable time-step errors are observed in $E_b$ for $\tau\le 0.05$, and the inset shows that the absolute energies are almost converged for $\tau\le 0.03$~a.u.
A value of $\tau=0.05$ would be enough for having no sizeable finite-time-step bias on the $E_b$ evaluations on the selected system, however in this work we are going to consider also larger 2D ice structures (up to 192 waters for supercell). 
Thus, we took for the production calculations a safe value of $\tau=0.02$~a.u., that is 10 times more efficient than the values that was considered for instance in \citet{Ice:prl2011} with the old DMC algorithm.

The second aspect that needs to be carefully checked in order to provide correct benchmark values are finite-size (FS) effects.
In contrast to DFT, which is an effective one-particle method that can exploit Bloch's theorem in calculations for extended periodic systems, DMC is a many-body method and the finite size errors can be sizeable unless large supercells and correction terms are considered~\cite{Lin:qmctwistavg:pre2001,Chiesa:size_effects:prl2006,KZK:prl2008}.
To eliminate the finite size errors, we performed several calculations for increasing supercell sizes up to \textit{ca.} 200 water molecules and the binding energy is extrapolated to the infinite size, as shown in Fig.~2 of the main manuscript.
Larger systems yield smaller FS error, but the computational effort increases with $N_w$ roughly linearly. 
The reason for having $E_b$ scaling as ${N_w}^{\alpha}$, with $\alpha \sim 1$, is because in DMC the stochastic error $\sigma$ is proportional to the inverse  of the square root of the computational cost, and, given a target stochastic error $\sigma$ on the total energy, an evaluation for a system with $N$ electrons has a computational cost $\propto N^\gamma$, with $\gamma \sim 3$.  Each water has 8 valence electrons (we use pseudo potentials), so $N=8 N_w$. Here, our target stochastic error is on the energy per water, thus we need an error $\propto N_w \sigma$, yielding $\alpha = \gamma -2$. However, the actual scaling is slightly worse than linear, because in DMC there is an equilibration time that gets removed; in small systems this equilibration time is a negligible fraction of the total computation, but it scales as $N^\gamma$ so it becomes relevant as the system size increases. Thanks to the algorithmic improvements of Ref.~\onlinecite{sizeconsDMC}, we can use large time-steps such that also in the huge systems here considered the equilibration time is not a bottleneck. 

\begin{figure*}[htbp]
\begin{center}
\includegraphics[width=5.in]{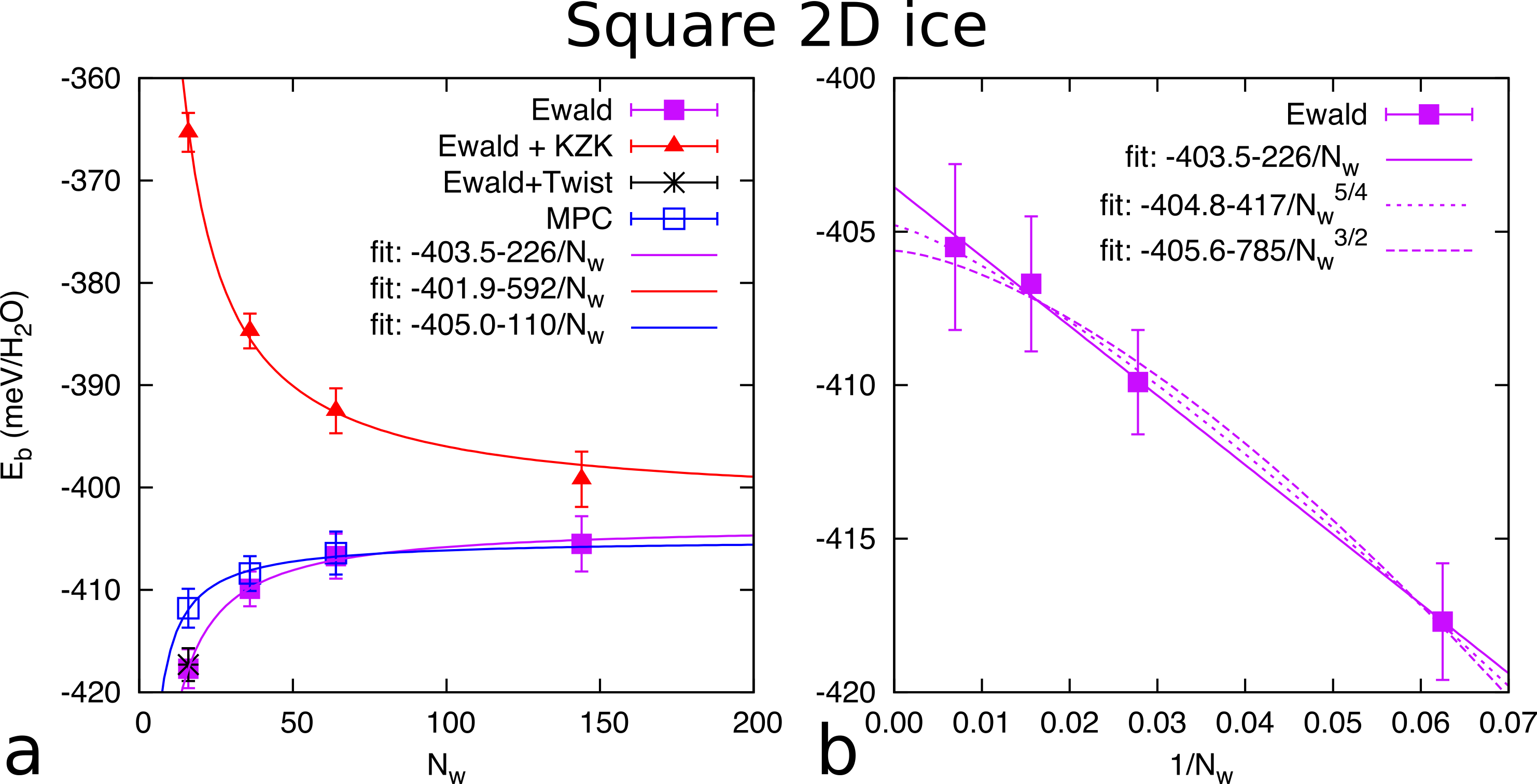}
\end{center}
\caption{
(panel a)
Binding energy $E_b$ of square ice, evaluated using DMC with time-step $\tau=0.02$~a.u., plotted as a function of the number $N_w$ of waters per simulated supercell. 
Error bars correspond to one standard deviation in the energy evaluation.
Different methods to account for the periodicity are tested here: Ewald 2D periodicity (violet square), Ewald plus KZK corrections (red triangle), Ewald periodicity and twist average over 30 randomly selected phases (black asterisk), MPC periodicity evaluated on top of Ewald sampling (empty blue square).
Corresponding fitting lines are also plotted, and the functional shape of the fitting function assume finite-size errors are $\propto 1/N_w$. 
(panel b) 
Same as panel a, but plotted as a function of $1/N_w$. For simplicity, only evaluations that used Ewald 2D periodicity are considered.
Three different fitting functions are considered, based on the assumption that the FS errors are: 
$\propto {N_w}^{-1}$, as done in panel a and in the letter (continuous line); 
$\propto {N_w}^{-5/4}$, according to what suggested by \citet{FSEqmc:PRB2008} (dotted line);
and 
$\propto {N_w}^{-3/2}$, according to what suggested by \citet{Ceperley:PRB1978} (dashed line); 
}
\label{fig:qmc:FS}
\end{figure*}

Here we  further discuss the FS effects and corrections using the model periodic Coulomb interaction~\cite{MPC:Fraser1996,MPC:Will1997,MPC:Kent1999} or the Kwee, Zhang and Krakauer  method~\cite{KZK:prl2008}.
Again, we performed our tests on the square 2D ice structure, this time keeping $\tau=0.02$~a.u. and considering different sizes of the supercell, namely by considering an increasing number $N_w$ of water molecules in the simulated supercell in order to observe the convergence of $E_b$.
Fig.~\ref{fig:qmc:FS} shows the results of these tests. 
Panel a leads to several interesting observations.
First, we observe that the black asterisk in the plot, representing a twist average DMC evaluation, is superimposed with the purple square, that is the evaluation at $\Gamma$. 
Thus, the twist average technique is not necessary here, as expected since the system is an insulator and the smallest considered supercell is already pretty large (it comprises 16 water molecules); we do not use twist average for larger supercells.
If the Kwee, Zhang and Krakauer~\cite{KZK:prl2008} (KZK) corrections are added on top of the Ewald evaluations, we observe that the FS errors are overcorrected, probably because of the difference between these systems and the homogeneous electron gas where KZK corrections were tuned.
Thus, bare Ewald evaluations are to be preferred here.
We also tested the use of the model periodic Coulomb (MPC) interaction~\cite{MPC:Fraser1996,MPC:Will1997,MPC:Kent1999} in place of the Ewald one (actually, in the reported MPC calculation the DMC propagation is performed using Ewald, as recommended in CASINO's manual, because there is some evidence that the MPC interaction can distort the XC hole).
MPC interaction mitigates the FS effects, although they are not completely removed and some sort of extrapolation to the infinite system is still necessary.
We have tested different possible extrapolations for the bare DMC evaluations with Ewald interaction, and they are reported in Fig.~\ref{fig:qmc:FS}b.
It is not know \textit{a priori} what the functional form of the FS bias is for a quasi-2D system as the one we are dealing now; for a bulk system the FS error is often expected to be $\propto {N_w}^{-1}$, and for strict 2D systems \citet{Ceperley:PRB1978}  proposed that they are $\propto {N_w}^{-3/2}$, whereas \citet{FSEqmc:PRB2008} proposed $\propto {N_w}^{-5/4}$.
We are not able to determine which is the best fitting function because the error bars of the DMC evaluation are much larger than the residue of the fitting. 
This is also because our smallest system is already very large: one or more orders of magnitude larger than what has been considered in other studies where infinite size extrapolations have been performed \cite{FSEqmc:PRB2008}.
Thus, we are much less affected here by the specific form chosen for the fitting function.
Indeed, we obtain no significant different in the extrapolation of $E_b$ in the three methods; the maximum difference in 2.1~meV, that is small enough for the accuracy that we need for our evaluations. 
Thus, in the letter we took the easiest choice for the fitting, that is a FS error $\propto {N_w}^{-1}$, upon the bare DMC evaluations with Ewald 2D periodicity.

The structures used for DMC calculations are obtained from relaxation using the optPBE-vdW-DF functional; 
The hexagonal structure is not considered in the case of 2~GPa pressure because its area per water $A$ is to much larger that for the other structures, thus the $PV=P\times A\times w$ term in the entropy makes this structure unstable already at much smaller pressures.


\subsection{DFT}
DFT calculations were performed using the Vienna Ab-initio Simulation Package (VASP), using hard projector augmented wave (PAW) potentials and a 1000 eV
plane wave cutoff.
A periodic slab model was used containing a vacum region and the cell dimension perpendicular
to the ice layer is 15 \AA.
The cell dimension perpendicular to the slab was fixed during structure optimization,
and the lateral cell dimensions were relaxed until the 
lateral diagonal elements of the stress tensor were 
converged to the target lateral pressures.
K-points were sampled using the Monkhorst-Pack grids, with $4 \times 4 \times 1$ 
for the hexagonal and pentagonal structures and $6 \times 6 \times 1$ for the 
square and rhombic structures.
Various DFT functionals of LDA \cite{perdew_self_1981}, GGA, meta-GGA and hybrid type were used.
GGA functionals used include the Perdew-Burke-Ernzerhof (PBE) \cite{perdew_generalized_1996},
revPBE \cite{zhang_comment_1998} and Becke-Lee-Yang-Parr (BLYP) \cite{PhysRevA.38.3098, PhysRevB.37.785}.
SCAN is a meta-GGA functional \cite{sun_accurate_2016}.
Hybrid functionals considered are
the PBE0 \cite{adamo_toward_1999} and HSE \cite{heyd_hybrid_2003, heyd_erratum:_2006} functionals.
Van der Waals corrections were accounted in several different schemes.
The first is the D2 and D3 corrections of Grimme \cite{grimme_semiempirical_2006, grimme_a_2010},
and superscript atm indicates a three-body dispersion term of Axilrod-Teller-Muto type.
The second is the corrections of Tkatchenko and Scheffler (+vdW(TS)) \cite{tkatchenko_accurate_2009}. 
The self consitent screening (scs) and 
many body dispersion (MBD) \cite{PhysRevLett.108.236402} were also considered with the TS scheme. %
The third contains the van der Waals functionals of Langreth and Lundqvist, including
PBE-vdW-DF, optPBE-vdW-DF, revPBE-vdW-DF, optB88-vdW-DF, optB86b-vdW-DF  \cite{dion_van_2004, roman-perez_efficient_2009, klimes_chemical_2010, klimes_van_2011} and the dW-DF2 functional (rPW86-vdW-DF2) \cite{lee_higher-accuracy_2010}.

\subsection{Low-cost electronic structure methods}
Three low-cost electronic structure methods has been considered.
A hybrid functional evaluated in a double-$\zeta$ type basis set HSE-3c~\cite{hse3c},
a minimal basis set Hartree-Fock method HF-3c~\cite{hf3c}, and a density functional tight-binding Hamiltonian
with re-parametrized dispersion correction DFTB3-D3~\cite{dftb3d3}.
The setups can be found in Ref. \onlinecite{Brandenburg_ice:2015}.
CRYSTAL14~\cite{crystal14} (develop version) was used to perform HSE-3c and HF-3c calculations.
DFTB+~\cite{dftbp} was used to perform DFTB3-D3 calculations.

\subsection{Force field}
Force field calculations of the SPC/E \cite{berendsen_missing_1987}, TIP4P \cite{jorgensen_comparison_1983}, TIP4P/2005 \cite{abascal_general_2005}, TIP4P/Ice \cite{abascal_a_2005} and coarse grained mW model \cite{molinero_water_2009}
were carried out using the LAMMPS code \cite{plimpton_fast_1995}.
A $10 \times 10 \times 1$ supercell was used and periodic boundary conditions were applied in 
the two lateral dimensions.
The
long-range Coulombic interaction is calculated using a particle-particle particle-mesh
(PPPM) algorithm with an accuracy of $10^{-4}$.
Energy minimization was performed with fixed cell dimensions and 
the lateral cell dimensions were
tuned manually until the total energy is converged to within 1 meV/$\text{H}_2\text{O}$.


\section{Extended data}
\label{sec:extdat}

\begin{figure}[h!]
\begin{center}
\includegraphics[width=3.3in]{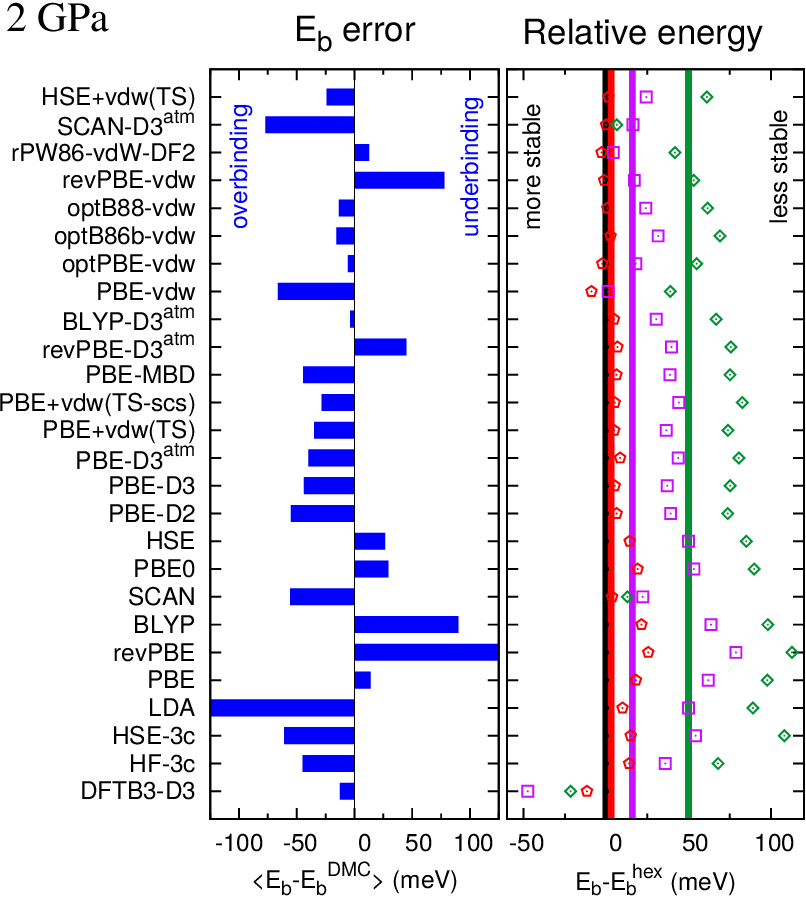}
\end{center}
\caption{ 
Left panel reports the average error (blue bars) and average absolute error (gray bars) for the $E_b$ evaluations, taking DMC as reference, using several electronic structure methods, covering low-cost approaches and DFT with LDA, GGA, meta-GGA, gybrid,  GGA+vdW, meta-GGA+vdW and hybrid+vdW functionals. 
Right panel reports the energy difference at 2~GPa pressure for each of the considered methods, and the  benchmark DMC evaluations are shown as color-coded bands, and their width represents the stochastic error (i.e., $\pm \sigma$). Color and symbol conventions are those of the figures in the letter.
Results are on the optPBE-vdW-DF structures at 2~GPa pressure.
}
\label{fig:2GPa}
\end{figure}

\begin{figure}[h!]
\begin{center}
\includegraphics[width=3.4in]{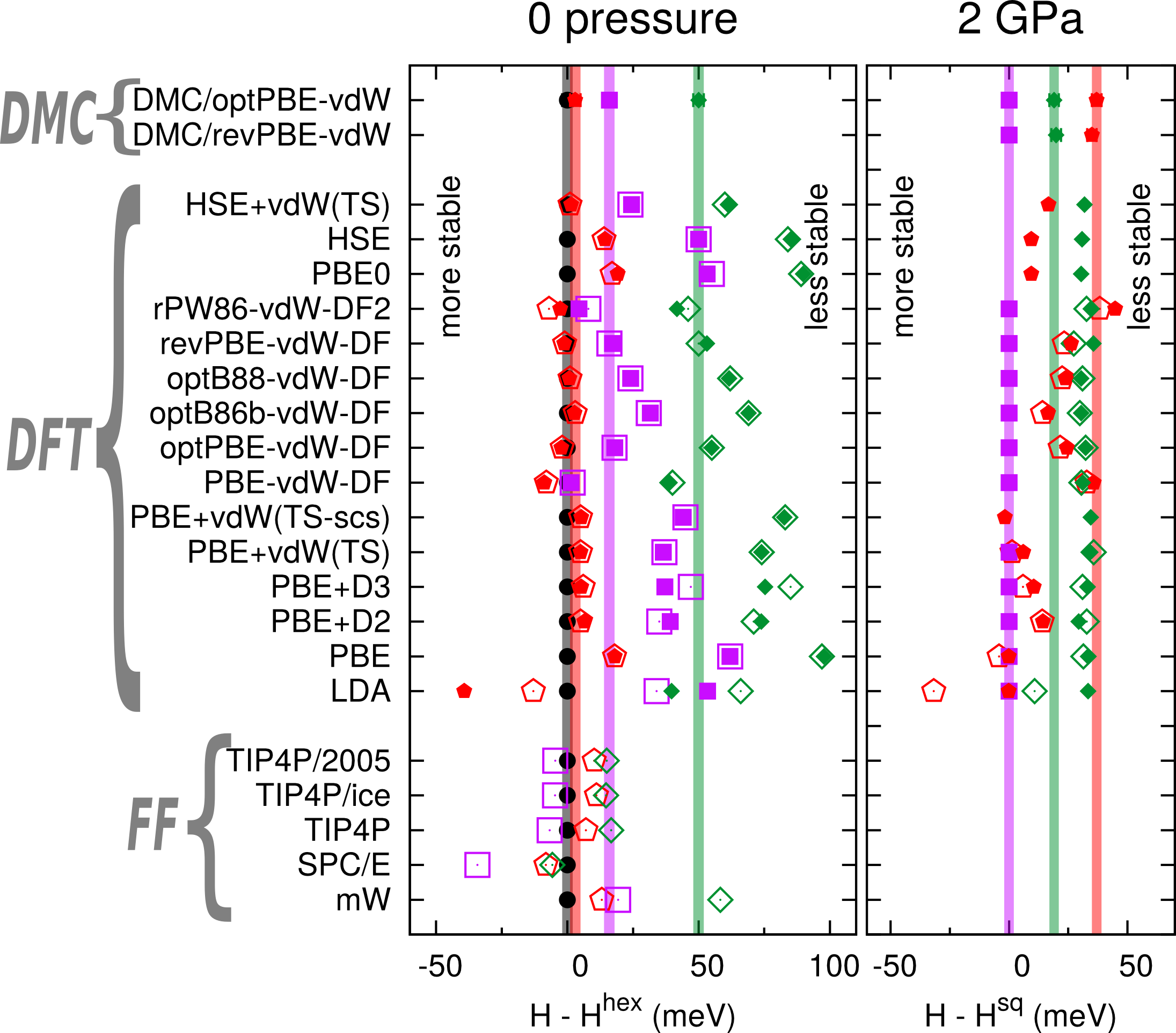}
\end{center}
\caption{ 
Enthalpy difference $\Delta H$, both at 0 pressure and at 2 GPa, of each 2D ice structure with respect to the most stable structure at the corresponding pressure (i.e., hexagonal at 0 P and square at 2 GPa), obtained by DMC, several different XC-functionals (covering LDA, GGA, GGA+vdW, hybrids and hybrids+vdW) and FF approaches. 
Nuclear quantum effects on $H$ are neglected. 
Color and symbol conventions are those of the figures in the letter.
The small filled symbols correspond to the results on the optPBE-vdW-DF structure, whereas the empty  symbols are for the relaxed configurations. 
For DMC we do not have DMC relaxed figures, but for the 2~GPa case we also show the results for the revPBE-vdW structures.
The color-coded bands represent the benchmark DMC/optPBE-vdW evaluations. 
}
\label{fig:geo}
\end{figure}

\begin{figure}[h!]
\begin{center}
\includegraphics[width=3.3in]{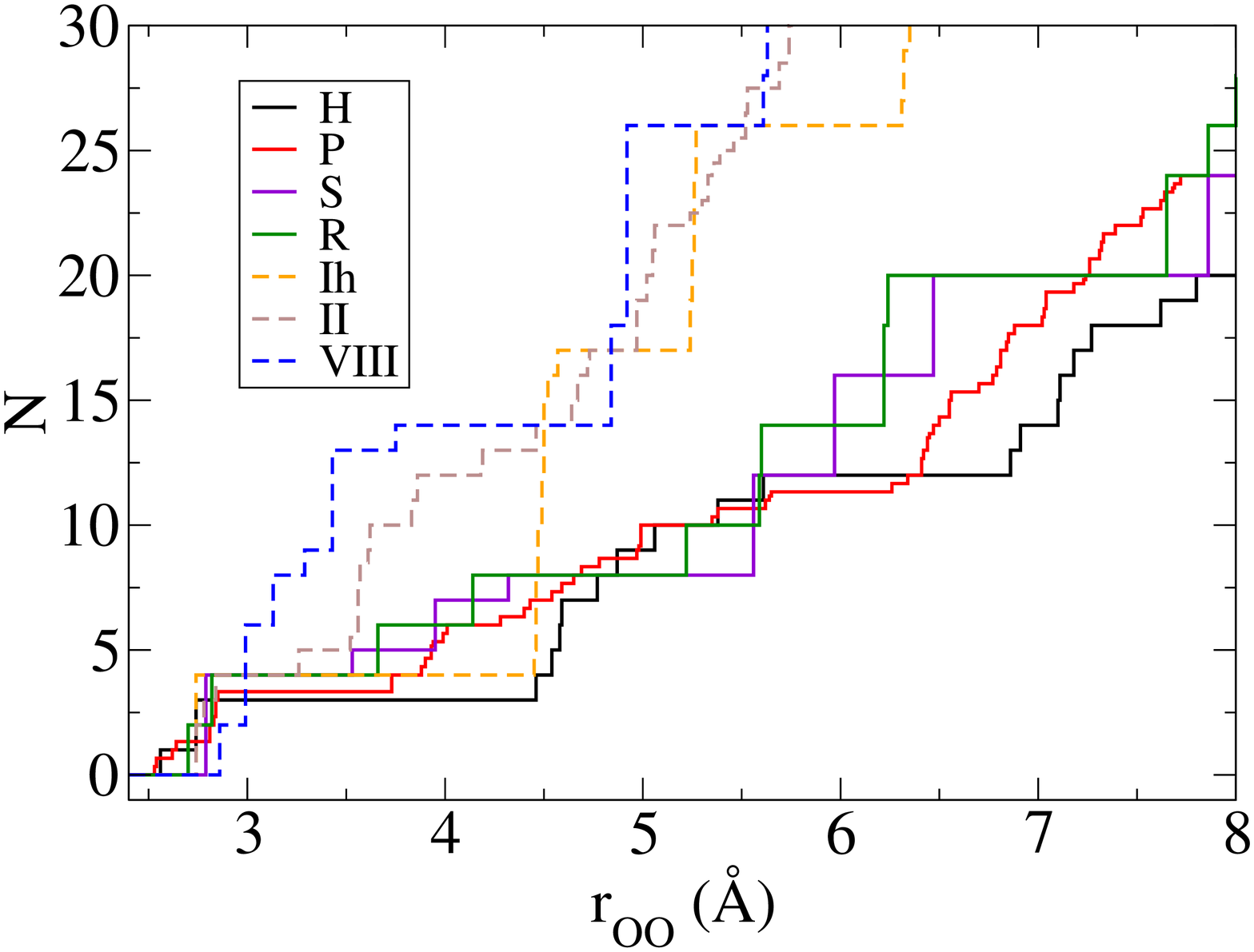}
\end{center}
\caption{
Integrated number of neighbors as a function of O-O distance for 2D (H, P, S, R) and bulk (Ih, II, VIII) structures at 0 pressure.
}
\label{fig_tip4p_N}
\end{figure}

\begin{figure}[h!]
\begin{center}
\includegraphics[width=3.3in]{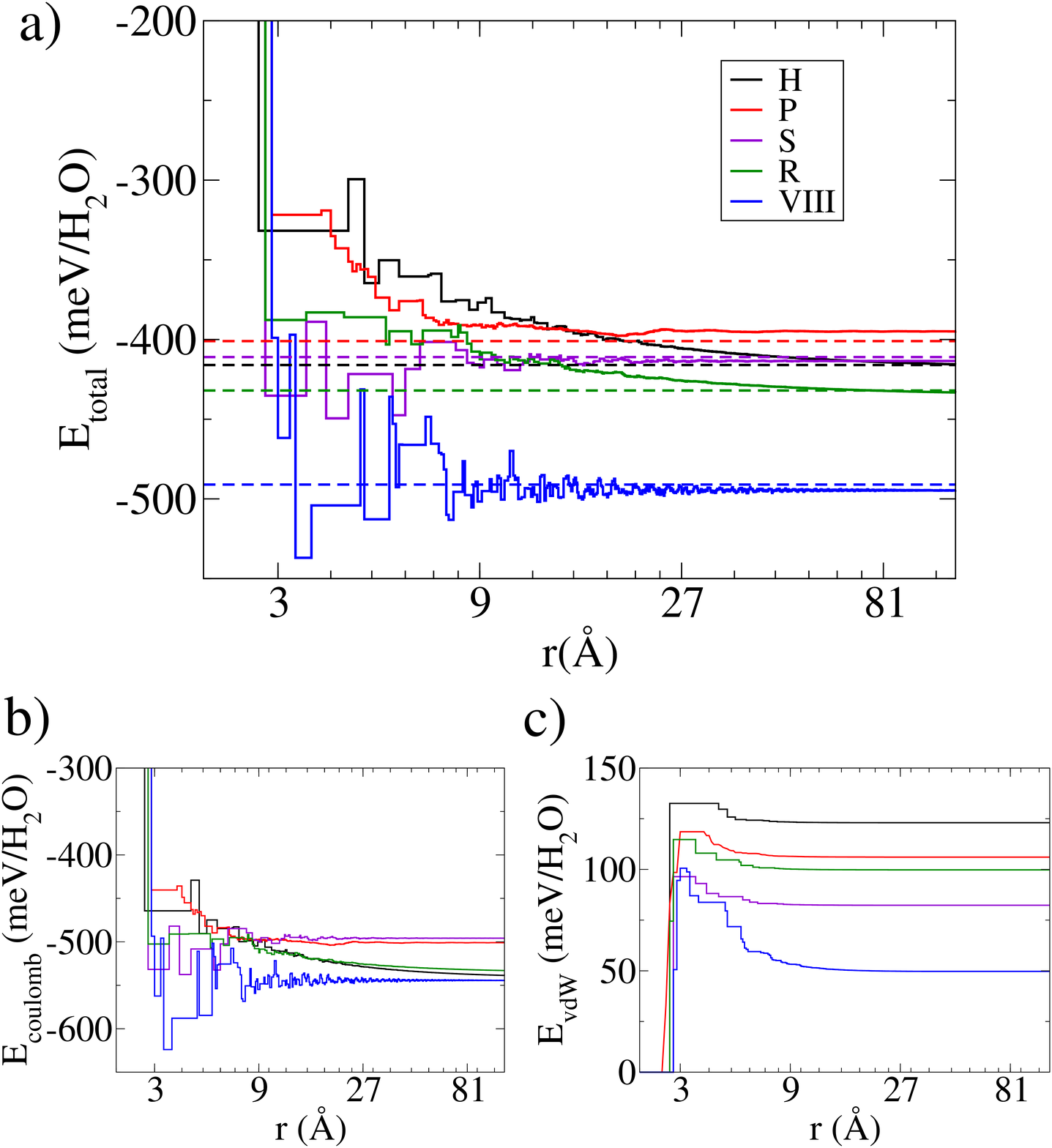}
\end{center}
\caption{
Total interaction (left), Coulomb interaction (middle) and van der Waals interaction (right) of TIP4P model as a function
of O-O distance for ice structures at 0 pressure.
On the left panel, the converged value calculated using the PPPM algorithm in LAMMPS.
}
\label{fig_tip4p_int}
\end{figure}

Benchmark on the high pressure structures is shown in Fig.~\ref{fig:2GPa}.
The effect of geometry relaxation on the enthalpy is shown in Fig.~\ref{fig:geo}.
Binding energy values calculated with different methods are reported in Table~\ref{table1}.
A comparison between the integrated number of neighbors as a function of the O-O distance for the 2D ice structures and the bulk ice (Ih, II, VIII) is shown in Fig.~\ref{fig_tip4p_N}, and in Fig.~\ref{fig_tip4p_int} we show the contribution in the interaction given by waters within a given  distance.

%
%
%


\begin{table*}[htdp]
\resizebox{16cm}{!}{
 \begin{tabular}{|c|c|c|c|c|c|c|c|}
  \hline
   $E_\text{b}$ (meV)  & H (0 GPa)   &  P (0 GPa) &  S (0 GPa) & R (0 GPa) & P (2 GPa) & S (2 GPa) & R (2 GPa) \\
  \hline
	DMC		& -423	& -419	& -404	& -389	& -380	& -374	& -385 \\
	HSE+vdw(TS) 	& -451	& -449	& -424	& -406	& -429 	& -403 	& -401 \\
	HSE 		& -417	& -401	& -364	& -347 	& -374 	& -341 	& -340 \\
	PBE0 		& -417 	& -397	& -361	& -343 	& -371 	& -338 	& -337 \\
	$\text{SCAN-D3}^\text{atm}$	& -489 	& -488 	& -469 	& -498 	& -488	& -478	& -480 \\	
	SCAN		& -471	& -467	& -446	& -474	& -466	& -452 	& -455 \\
	$\text{BLYP-D3}^\text{atm}$	& -435	& -429	& -402	& -384	& -397	& -375	& -372 \\
	BLYP		& -362	& -339	& -295	& -280	& -291	& -253	& -247 \\
	rPW86-vdW-DF2 	& -404	& -406	& -397	& -378	& -346	& -348	& -344 \\
	optB88-vdw 	& -441	& -439	& -414	& -395	& -417	& -398	& -398 \\
	optB86b-vdw 	& -447	& -444	& -413	& -394	& -425	& -399	& -398 \\
	optPBE-vdw 	& -429	& -430	& -408	& -390	& -392	& -374	& -371 \\
	revPBE-vdw 	& -345	& -345	& -325	& -308	& -280	& -263	& -258 \\
	$\text{revPBE-D3}^{\text{atm}}$	& -391	& -384	& -349	& -332	& -348	& -316	& -311 \\
	revPBE 		& -303	& -276	& -221	& -206	& -220	& -166	& -158 \\
	PBE-vdw 	& -480	& -488	& -476	& -457	& -453	& -446	& -445 \\
	PBE-MBD		& -480	& -473	& -439	& -421	& -457	& -425	& -424 \\
	PBE+vdw(TS-scs) & -467	& -461	& -420	& -401	& -440	& -395	& -391 \\
	PBE+vdw(TS) 	& -469	& -464	& -430	& -412	& -446	& -409	& -405 \\
	$\text{PBE-D3}^\text{atm}$	& -479	& -469	& -432	& -414	& -453	& -418	& -416 \\
	PBE-D3  	& -479	& -473	& -439	& -420	& -456	& -424	& -421 \\
	PBE-D2  	& -491	& -483	& -449	& -433	& -473	& -445	& -445 \\
	PBE  		& -436	& -417	& -371	& -355	& -393	& -350	& -347 \\
	LDA  		& -708	& -697	& -655	& -635 	& -779	& -713 	& -774 \\
	HF-3c		& -480	& -465	& -441	& -428	& -438	& -420	& -408 \\
	HSE-3c		& -511 	& -495	& -454	& -419	& -474	& -430	& -395 \\
	DFTB3-D3	& -398	& -409	& -443	& -436	& -396	& -455	& -457 \\
  \hline
 \end{tabular}
}
 \caption{
Binding energies (Eq. \ref{eq:Eb}) of
2D ice calculated using different methods.
}
\label{table1}
\end{table*}

\begin{table}[htdp]
\begin{center}
\begin{tabular}{ | l |  c c  |  c c | }
\hline
                &    optPBE-vdW     &                     & revPBE-vdW        &   \\  
                &   $\Delta_\text{sq}$  & $\Delta_\text{sq}^\text{300K}$ &    $\Delta_\text{sq}$  & $\Delta_\text{sq}^\text{300K}$ \\
\hline         
Pentagonal      &  \textbf{28}      &       \textbf{14}   &  \textbf{34}      &       \textbf{32}  \\
Square          &  \textbf{0}       &       \textbf{0}    &  \textbf{0}       &       \textbf{0}  \\
Rhombic         &  \textbf{17}      &       \textbf{18}   &  \textbf{16}      &       \textbf{29}  \\
\hline
\end{tabular}
\caption{
Enthalpy and free energy difference (meV/H$_2$O) at 2 GPa.
See the main text for the definition of $\Delta_\text{sq}$.
$\Delta_\text{sq}^\text{300K}$ is the free energy difference where the vibrational free energy at 300 K is considered.
}
\end{center}
\end{table}

\section{Additional notes}
\label{sec:note}

Recently, a hierarchy of the low-cost electronic structure methods has been developed in the Grimme group 
(HSE-3c\cite{hse3c}, HF-3c\cite{hf3c}, and DFTB3-D3\cite{dftb3d3}).
They allow speed-ups of one, two, and four orders of magnitudes compared to the DFT calculations.
We find that 2D ice is still challenging and especially the tight-binding accuracy is not satisfactory.
This is probably due to
very approximate description of short-range exchange repulsion and induction based on monopole charges.
However, the accuracy of these methods is reasonable comparing to many DFT methods and might
be useful for screening applications.

\section{Structures}\label{sec:strfiles}

\begin{figure}[h!]
\begin{center}
\includegraphics[width=3.4in]{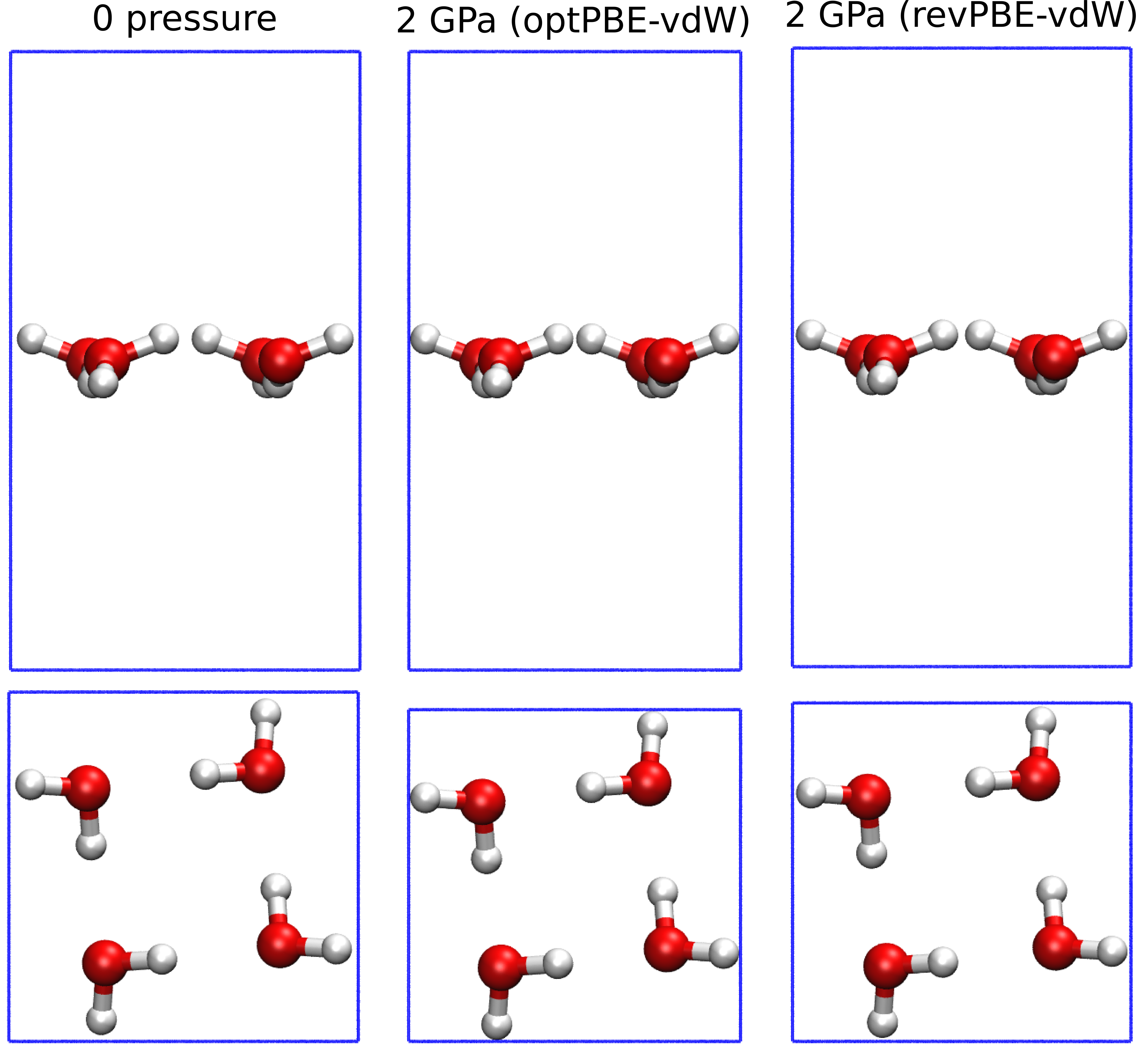}
\end{center}
\caption{
Side and top view of the Square structures used in the DMC calculations, obtained from DFT-based optimization for the zero pressure case and with the optPBE-vdW-DF functional (left), the 2~GPa pressure case and  optPBE-vdW-DF functional (middle), and the 2~GPa pressure case and  revPBE-vdW-DF functional (right). The blue boxes represent the primitive unit cell. 
}
\label{fig:Sgeo}
\end{figure}

\begin{figure}[h!]
\begin{center}
\includegraphics[width=3.4in]{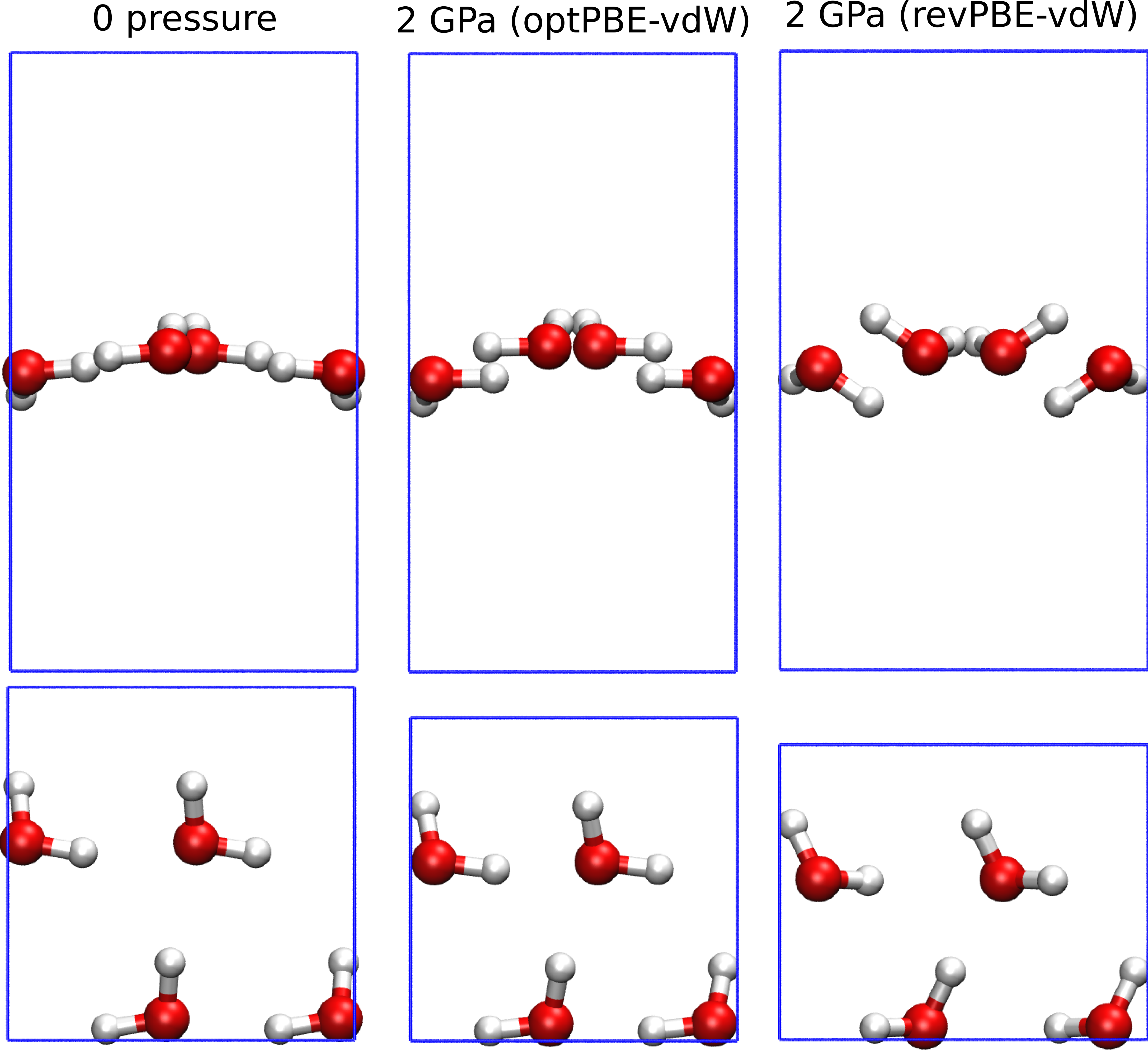}
\end{center}
\caption{
Same of Fig.~\ref{fig:Sgeo}, but for the rhombic structure.
}
\label{fig:Rgeo}
\end{figure}

2D ice structures used for DMC calculations with .cif file format.
(i - iv) Structures at 0 GPa calculated with optPBE-vdW functional.
(v - vii) Structures at 2 GPa calculated with optPBE-vdW functional.
(viii - x) Structures at 2 GPa calculated with revPBE-vdW functional.
The square and rhombic structures are also shown in Fig.~\ref{fig:Sgeo} and \ref{fig:Rgeo}, respectively.

\newpage
%

\newpage
\begin{widetext}
{\footnotesize 
\begin{verbatim}
# i. Structure H, Pressure 0, optPBE-vdW

data


_pd_phase_name                         'H   O                                 ' 
_cell_length_a                         9.35661 
_cell_length_b                         8.31717 
_cell_length_c                         15.00000 
_cell_angle_alpha                      90 
_cell_angle_beta                       90 
_cell_angle_gamma                      90 
_symmetry_space_group_name_H-M         'P 1' 
_symmetry_Int_Tables_number            1 

loop_ 
_symmetry_equiv_pos_as_xyz 
   'x, y, z' 

loop_ 
   _atom_site_label 
   _atom_site_occupancy 
   _atom_site_fract_x 
   _atom_site_fract_y 
   _atom_site_fract_z 
   _atom_site_adp_type 
   _atom_site_B_iso_or_equiv 
   _atom_site_type_symbol 
   H1         1.0     0.696420      0.546955      0.489865     Biso  1.000000 H 
   H2         1.0     0.613038      0.714406      0.491282     Biso  1.000000 H 
   H3         1.0     0.303580      0.046955      0.510135     Biso  1.000000 H 
   H4         1.0     0.386962      0.214406      0.508718     Biso  1.000000 H 
   H5         1.0     0.848566      0.423065      0.570672     Biso  1.000000 H 
   H6         1.0     0.953181      0.496056      0.500791     Biso  1.000000 H 
   H7         1.0     0.151434      0.923065      0.429328     Biso  1.000000 H 
   H8         1.0     0.046819      0.996056      0.499209     Biso  1.000000 H 
   H9         1.0     0.803579      0.046955      0.489865     Biso  1.000000 H 
   H10        1.0     0.886962      0.214406      0.491282     Biso  1.000000 H 
   H11        1.0     0.196421      0.546955      0.510135     Biso  1.000000 H 
   H12        1.0     0.113038      0.714406      0.508718     Biso  1.000000 H 
   H13        1.0     0.348566      0.423065      0.429328     Biso  1.000000 H 
   H14        1.0     0.453181      0.496055      0.499209     Biso  1.000000 H 
   H15        1.0     0.651434      0.923065      0.570672     Biso  1.000000 H 
   H16        1.0     0.546819      0.996055      0.500791     Biso  1.000000 H 
   O1         1.0     0.602125      0.597283      0.498958     Biso  1.000000 O 
   O2         1.0     0.397875      0.097283      0.501042     Biso  1.000000 O 
   O3         1.0     0.862152      0.431677      0.506394     Biso  1.000000 O 
   O4         1.0     0.137848      0.931677      0.493606     Biso  1.000000 O 
   O5         1.0     0.897875      0.097283      0.498958     Biso  1.000000 O 
   O6         1.0     0.102125      0.597283      0.501042     Biso  1.000000 O 
   O7         1.0     0.362152      0.431677      0.493606     Biso  1.000000 O 
   O8         1.0     0.637848      0.931677      0.506394     Biso  1.000000 O 

#====================================================================== 

# ii. Structure P, Pressure 0, optPBE-vdW

data


_pd_phase_name                         'H  O                                  ' 
_cell_length_a                         10.18521 
_cell_length_b                         10.17328 
_cell_length_c                         15.00000 
_cell_angle_alpha                      90 
_cell_angle_beta                       90 
_cell_angle_gamma                      89.99858 
_symmetry_space_group_name_H-M         'P 1' 
_symmetry_Int_Tables_number            1 

loop_ 
_symmetry_equiv_pos_as_xyz 
   'x, y, z' 

loop_ 
   _atom_site_label 
   _atom_site_occupancy 
   _atom_site_fract_x 
   _atom_site_fract_y 
   _atom_site_fract_z 
   _atom_site_adp_type 
   _atom_site_B_iso_or_equiv 
   _atom_site_type_symbol 
   H1         1.0     0.058799      0.175119      0.492877     Biso  1.000000 H 
   H2         1.0     0.903013      0.170871      0.503135     Biso  1.000000 H 
   H3         1.0     0.991017      0.950631      0.500095     Biso  1.000000 H 
   H4         1.0     0.066596      0.818289      0.524854     Biso  1.000000 H 
   H5         1.0     0.164316      0.579537      0.500567     Biso  1.000000 H 
   H6         1.0     0.165822      0.426958      0.478815     Biso  1.000000 H 
   H7         1.0     0.307474      0.925503      0.509516     Biso  1.000000 H 
   H8         1.0     0.441846      0.004735      0.501498     Biso  1.000000 H 
   H9         1.0     0.558825      0.325374      0.507073     Biso  1.000000 H 
   H10        1.0     0.403043      0.329653      0.496819     Biso  1.000000 H 
   H11        1.0     0.566629      0.682256      0.475148     Biso  1.000000 H 
   H12        1.0     0.491062      0.549918      0.499918     Biso  1.000000 H 
   H13        1.0     0.664346      0.920985      0.499399     Biso  1.000000 H 
   H14        1.0     0.665860      0.073576      0.521123     Biso  1.000000 H 
   H15        1.0     0.807406      0.574996      0.490467     Biso  1.000000 H 
   H16        1.0     0.941777      0.495756      0.498452     Biso  1.000000 H 
   H17        1.0     0.225340      0.259312      0.570005     Biso  1.000000 H 
   H18        1.0     0.275775      0.153118      0.501781     Biso  1.000000 H 
   H19        1.0     0.342873      0.709516      0.496296     Biso  1.000000 H 
   H20        1.0     0.232407      0.759296      0.430445     Biso  1.000000 H 
   H21        1.0     0.725344      0.241163      0.429893     Biso  1.000000 H 
   H22        1.0     0.775753      0.347372      0.498121     Biso  1.000000 H 
   H23        1.0     0.842873      0.790978      0.503705     Biso  1.000000 H 
   H24        1.0     0.732399      0.741193      0.569549     Biso  1.000000 H 
   O1         1.0     0.983052      0.116981      0.501413     Biso  1.000000 O 
   O2         1.0     0.987080      0.852861      0.497333     Biso  1.000000 O 
   O3         1.0     0.111422      0.498554      0.501391     Biso  1.000000 O 
   O4         1.0     0.344806      0.012968      0.499354     Biso  1.000000 O 
   O5         1.0     0.483076      0.383541      0.498622     Biso  1.000000 O 
   O6         1.0     0.487123      0.647686      0.502682     Biso  1.000000 O 
   O7         1.0     0.611468      0.001980      0.498540     Biso  1.000000 O 
   O8         1.0     0.844738      0.487518      0.500589     Biso  1.000000 O 
   O9         1.0     0.236994      0.244865      0.506185     Biso  1.000000 O 
   O10        1.0     0.249322      0.743729      0.493599     Biso  1.000000 O 
   O11        1.0     0.736995      0.255614      0.493712     Biso  1.000000 O 
   O12        1.0     0.749331      0.756741      0.506395     Biso  1.000000 O 

#====================================================================== 

# iii. Structure S, Pressure 0, optPBE-vdW

data


_pd_phase_name                         'H   O                                 ' 
_cell_length_a                         5.64749 
_cell_length_b                         5.64746 
_cell_length_c                         15.00000 
_cell_angle_alpha                      90 
_cell_angle_beta                       90 
_cell_angle_gamma                      90 
_symmetry_space_group_name_H-M         'P 1' 
_symmetry_Int_Tables_number            1 

loop_ 
_symmetry_equiv_pos_as_xyz 
   'x, y, z' 

loop_ 
   _atom_site_label 
   _atom_site_occupancy 
   _atom_site_fract_x 
   _atom_site_fract_y 
   _atom_site_fract_z 
   _atom_site_adp_type 
   _atom_site_B_iso_or_equiv 
   _atom_site_type_symbol 
   H1         1.0     0.562438      0.764627      0.523666     Biso  1.000000 H 
   H2         1.0     0.735373      0.937563      0.476320     Biso  1.000000 H 
   H3         1.0     0.437562      0.235373      0.523666     Biso  1.000000 H 
   H4         1.0     0.264627      0.062438      0.476320     Biso  1.000000 H 
   H5         1.0     0.062438      0.735373      0.523666     Biso  1.000000 H 
   H6         1.0     0.235373      0.562437      0.476320     Biso  1.000000 H 
   H7         1.0     0.937562      0.264627      0.523666     Biso  1.000000 H 
   H8         1.0     0.764627      0.437563      0.476320     Biso  1.000000 H 
   O1         1.0     0.724252      0.775748      0.499993     Biso  1.000000 O 
   O2         1.0     0.275748      0.224252      0.499993     Biso  1.000000 O 
   O3         1.0     0.224252      0.724252      0.499993     Biso  1.000000 O 
   O4         1.0     0.775748      0.275748      0.499993     Biso  1.000000 O 

#======================================================================

# iv. Structure R, Pressure 0, optPBE-vdW

data


_pd_phase_name                         'H O                                   ' 
_cell_length_a                         5.60546 
_cell_length_b                         5.70809 
_cell_length_c                         20.00000 
_cell_angle_alpha                      90 
_cell_angle_beta                       90 
_cell_angle_gamma                      90.00002 
_symmetry_space_group_name_H-M         'P 1' 
_symmetry_Int_Tables_number            1 

loop_ 
_symmetry_equiv_pos_as_xyz 
   'x, y, z' 

loop_ 
   _atom_site_label 
   _atom_site_occupancy 
   _atom_site_fract_x 
   _atom_site_fract_y 
   _atom_site_fract_z 
   _atom_site_adp_type 
   _atom_site_B_iso_or_equiv 
   _atom_site_type_symbol 
   H1         1.0     0.468154      0.219348      0.527052     Biso  1.000000 H 
   H2         1.0     0.284678      0.031820      0.504323     Biso  1.000000 H 
   H3         1.0     0.968153      0.219346      0.472946     Biso  1.000000 H 
   H4         1.0     0.784678      0.031820      0.495674     Biso  1.000000 H 
   H5         1.0     0.031789      0.719346      0.472946     Biso  1.000000 H 
   H6         1.0     0.215265      0.531819      0.495674     Biso  1.000000 H 
   H7         1.0     0.715265      0.531819      0.504323     Biso  1.000000 H 
   H8         1.0     0.531789      0.719347      0.527052     Biso  1.000000 H 
   O1         1.0     0.456895      0.060563      0.508584     Biso  1.000000 O 
   O2         1.0     0.956895      0.060561      0.491413     Biso  1.000000 O 
   O3         1.0     0.043048      0.560562      0.491413     Biso  1.000000 O 
   O4         1.0     0.543048      0.560562      0.508584     Biso  1.000000 O 

#======================================================================

# v. Structure P, Pressure 2 GPa, optPBE-vdW

data


_pd_phase_name                         'H  O                                  ' 
_cell_length_a                         9.64283 
_cell_length_b                         9.59451 
_cell_length_c                         15.00000 
_cell_angle_alpha                      90 
_cell_angle_beta                       90 
_cell_angle_gamma                      90.00531 
_symmetry_space_group_name_H-M         'P 1' 
_symmetry_Int_Tables_number            1 

loop_ 
_symmetry_equiv_pos_as_xyz 
   'x, y, z' 

loop_ 
   _atom_site_label 
   _atom_site_occupancy 
   _atom_site_fract_x 
   _atom_site_fract_y 
   _atom_site_fract_z 
   _atom_site_adp_type 
   _atom_site_B_iso_or_equiv 
   _atom_site_type_symbol 
   H1         1.0     0.059025      0.181096      0.493680     Biso  1.000000 H 
   H2         1.0     0.892657      0.171883      0.501419     Biso  1.000000 H 
   H3         1.0     0.995653      0.954918      0.499700     Biso  1.000000 H 
   H4         1.0     0.077948      0.813438      0.525118     Biso  1.000000 H 
   H5         1.0     0.166855      0.584504      0.499613     Biso  1.000000 H 
   H6         1.0     0.164166      0.420630      0.480485     Biso  1.000000 H 
   H7         1.0     0.300146      0.919066      0.508578     Biso  1.000000 H 
   H8         1.0     0.442959      0.004560      0.501162     Biso  1.000000 H 
   H9         1.0     0.559029      0.319348      0.506184     Biso  1.000000 H 
   H10        1.0     0.392662      0.328616      0.498470     Biso  1.000000 H 
   H11        1.0     0.577720      0.686845      0.474521     Biso  1.000000 H 
   H12        1.0     0.495601      0.545432      0.500163     Biso  1.000000 H 
   H13        1.0     0.667006      0.915957      0.499978     Biso  1.000000 H 
   H14        1.0     0.664333      0.079759      0.519365     Biso  1.000000 H 
   H15        1.0     0.800114      0.581679      0.490955     Biso  1.000000 H 
   H16        1.0     0.942874      0.496122      0.498533     Biso  1.000000 H 
   H17        1.0     0.218109      0.263405      0.572771     Biso  1.000000 H 
   H18        1.0     0.272985      0.147486      0.504530     Biso  1.000000 H 
   H19        1.0     0.355286      0.707297      0.494282     Biso  1.000000 H 
   H20        1.0     0.233592      0.758404      0.428558     Biso  1.000000 H 
   H21        1.0     0.718106      0.237122      0.427190     Biso  1.000000 H 
   H22        1.0     0.772945      0.353158      0.495317     Biso  1.000000 H 
   H23        1.0     0.855355      0.793144      0.505356     Biso  1.000000 H 
   H24        1.0     0.733367      0.742187      0.570893     Biso  1.000000 H 
   O1         1.0     0.979768      0.118133      0.502068     Biso  1.000000 O 
   O2         1.0     0.995380      0.850170      0.495894     Biso  1.000000 O 
   O3         1.0     0.109073      0.499545      0.501419     Biso  1.000000 O 
   O4         1.0     0.339502      0.012226      0.498796     Biso  1.000000 O 
   O5         1.0     0.479830      0.382273      0.497497     Biso  1.000000 O 
   O6         1.0     0.495346      0.650186      0.504009     Biso  1.000000 O 
   O7         1.0     0.609163      0.000884      0.498450     Biso  1.000000 O 
   O8         1.0     0.839410      0.488521      0.500815     Biso  1.000000 O 
   O9         1.0     0.231734      0.247461      0.509106     Biso  1.000000 O 
   O10        1.0     0.253240      0.741533      0.491478     Biso  1.000000 O 
   O11        1.0     0.731815      0.253104      0.490846     Biso  1.000000 O 
   O12        1.0     0.753247      0.759050      0.508006     Biso  1.000000 O 

#======================================================================

# vi. Structure S, Pressure 2 GPa, optPBE-vdW

data


_pd_phase_name                         'H   O                                 ' 
_cell_length_a                         5.36395 
_cell_length_b                         5.36359 
_cell_length_c                         20.00000 
_cell_angle_alpha                      90 
_cell_angle_beta                       90 
_cell_angle_gamma                      90 
_symmetry_space_group_name_H-M         'P 1' 
_symmetry_Int_Tables_number            1 

loop_ 
_symmetry_equiv_pos_as_xyz 
   'x, y, z' 

loop_ 
   _atom_site_label 
   _atom_site_occupancy 
   _atom_site_fract_x 
   _atom_site_fract_y 
   _atom_site_fract_z 
   _atom_site_adp_type 
   _atom_site_B_iso_or_equiv 
   _atom_site_type_symbol 
   H1         1.0     0.550905      0.765380      0.517722     Biso  1.000000 H 
   H2         1.0     0.734614      0.949058      0.482263     Biso  1.000000 H 
   H3         1.0     0.449095      0.234620      0.517722     Biso  1.000000 H 
   H4         1.0     0.265386      0.050942      0.482263     Biso  1.000000 H 
   H5         1.0     0.050905      0.734620      0.517722     Biso  1.000000 H 
   H6         1.0     0.234614      0.550942      0.482263     Biso  1.000000 H 
   H7         1.0     0.949095      0.265380      0.517722     Biso  1.000000 H 
   H8         1.0     0.765386      0.449058      0.482263     Biso  1.000000 H 
   O1         1.0     0.721789      0.778165      0.499993     Biso  1.000000 O 
   O2         1.0     0.278211      0.221835      0.499993     Biso  1.000000 O 
   O3         1.0     0.221789      0.721835      0.499993     Biso  1.000000 O 
   O4         1.0     0.778211      0.278165      0.499993     Biso  1.000000 O
#======================================================================

# vii. Structure R, Pressure 2 GPa, optPBE-vdW

data


_pd_phase_name                         'H O                                   ' 
_cell_length_a                         5.28492 
_cell_length_b                         5.22618 
_cell_length_c                         20.00000 
_cell_angle_alpha                      90 
_cell_angle_beta                       90 
_cell_angle_gamma                      89.99957 
_symmetry_space_group_name_H-M         'P 1' 
_symmetry_Int_Tables_number            1 

loop_ 
_symmetry_equiv_pos_as_xyz 
   'x, y, z' 

loop_ 
   _atom_site_label 
   _atom_site_occupancy 
   _atom_site_fract_x 
   _atom_site_fract_y 
   _atom_site_fract_z 
   _atom_site_adp_type 
   _atom_site_B_iso_or_equiv 
   _atom_site_type_symbol 
   H1         1.0     0.457729      0.226267      0.531500     Biso  1.000000 H 
   H2         1.0     0.242230      0.031526      0.512450     Biso  1.000000 H 
   H3         1.0     0.957705      0.226213      0.468488     Biso  1.000000 H 
   H4         1.0     0.742203      0.031509      0.487546     Biso  1.000000 H 
   H5         1.0     0.042209      0.726193      0.468479     Biso  1.000000 H 
   H6         1.0     0.257737      0.531535      0.487539     Biso  1.000000 H 
   H7         1.0     0.757731      0.531553      0.512462     Biso  1.000000 H 
   H8         1.0     0.542210      0.726235      0.531510     Biso  1.000000 H 
   O1         1.0     0.427204      0.053984      0.513267     Biso  1.000000 O 
   O2         1.0     0.927178      0.053964      0.486737     Biso  1.000000 O 
   O3         1.0     0.072763      0.553955      0.486733     Biso  1.000000 O 
   O4         1.0     0.572756      0.553980      0.513264     Biso  1.000000 O 

#======================================================================

# viii. Structure P, Pressure 2 GPa, revPBE-vdW

data


_pd_phase_name                         'H  O                                  ' 
_cell_length_a                         9.82443 
_cell_length_b                         9.76209 
_cell_length_c                         15.00000 
_cell_angle_alpha                      90 
_cell_angle_beta                       90 
_cell_angle_gamma                      89.99776 
_symmetry_space_group_name_H-M         'P 1' 
_symmetry_Int_Tables_number            1 

loop_ 
_symmetry_equiv_pos_as_xyz 
   'x, y, z' 

loop_ 
   _atom_site_label 
   _atom_site_occupancy 
   _atom_site_fract_x 
   _atom_site_fract_y 
   _atom_site_fract_z 
   _atom_site_adp_type 
   _atom_site_B_iso_or_equiv 
   _atom_site_type_symbol 
   H1         1.0     0.054960      0.180631      0.492126     Biso  1.000000 H 
   H2         1.0     0.893451      0.171204      0.502078     Biso  1.000000 H 
   H3         1.0     0.998560      0.952375      0.500611     Biso  1.000000 H 
   H4         1.0     0.076532      0.818046      0.533367     Biso  1.000000 H 
   H5         1.0     0.166631      0.582113      0.500697     Biso  1.000000 H 
   H6         1.0     0.163386      0.424322      0.476440     Biso  1.000000 H 
   H7         1.0     0.303581      0.918772      0.511649     Biso  1.000000 H 
   H8         1.0     0.440917      0.003450      0.501851     Biso  1.000000 H 
   H9         1.0     0.554911      0.319676      0.507680     Biso  1.000000 H 
   H10        1.0     0.393381      0.329110      0.497723     Biso  1.000000 H 
   H11        1.0     0.576581      0.682075      0.466349     Biso  1.000000 H 
   H12        1.0     0.498506      0.547872      0.499170     Biso  1.000000 H 
   H13        1.0     0.666469      0.918634      0.498883     Biso  1.000000 H 
   H14        1.0     0.663309      0.076356      0.523345     Biso  1.000000 H 
   H15        1.0     0.803603      0.581818      0.487995     Biso  1.000000 H 
   H16        1.0     0.941008      0.497317      0.497850     Biso  1.000000 H 
   H17        1.0     0.216461      0.263790      0.570909     Biso  1.000000 H 
   H18        1.0     0.270541      0.149800      0.504582     Biso  1.000000 H 
   H19        1.0     0.354429      0.706476      0.493231     Biso  1.000000 H 
   H20        1.0     0.232664      0.758905      0.431750     Biso  1.000000 H 
   H21        1.0     0.716660      0.236745      0.429012     Biso  1.000000 H 
   H22        1.0     0.770705      0.350784      0.495312     Biso  1.000000 H 
   H23        1.0     0.854143      0.793994      0.506421     Biso  1.000000 H 
   H24        1.0     0.732234      0.741729      0.567805     Biso  1.000000 H 
   O1         1.0     0.978378      0.118242      0.501637     Biso  1.000000 O 
   O2         1.0     0.000858      0.850328      0.497338     Biso  1.000000 O 
   O3         1.0     0.110126      0.498834      0.501765     Biso  1.000000 O 
   O4         1.0     0.340122      0.010106      0.499691     Biso  1.000000 O 
   O5         1.0     0.478337      0.382019      0.498003     Biso  1.000000 O 
   O6         1.0     0.500965      0.649915      0.502478     Biso  1.000000 O 
   O7         1.0     0.610037      0.001968      0.497880     Biso  1.000000 O 
   O8         1.0     0.840226      0.490558      0.500067     Biso  1.000000 O 
   O9         1.0     0.230910      0.246224      0.507642     Biso  1.000000 O 
   O10        1.0     0.255808      0.738171      0.493474     Biso  1.000000 O 
   O11        1.0     0.731121      0.254341      0.492277     Biso  1.000000 O 
   O12        1.0     0.755487      0.762442      0.506099     Biso  1.000000 O

#======================================================================

# ix. Structure S, Pressure 2 GPa, revPBE-vdW

data


_pd_phase_name                         'H   O                                 ' 
_cell_length_a                         5.47137 
_cell_length_b                         5.47131 
_cell_length_c                         20.00000 
_cell_angle_alpha                      90 
_cell_angle_beta                       90 
_cell_angle_gamma                      90 
_symmetry_space_group_name_H-M         'P 1' 
_symmetry_Int_Tables_number            1 

loop_ 
_symmetry_equiv_pos_as_xyz 
   'x, y, z' 

loop_ 
   _atom_site_label 
   _atom_site_occupancy 
   _atom_site_fract_x 
   _atom_site_fract_y 
   _atom_site_fract_z 
   _atom_site_adp_type 
   _atom_site_B_iso_or_equiv 
   _atom_site_type_symbol 
   H1         1.0     0.556014      0.766013      0.518061     Biso  1.000000 H 
   H2         1.0     0.733994      0.943984      0.481920     Biso  1.000000 H 
   H3         1.0     0.443986      0.233987      0.518061     Biso  1.000000 H 
   H4         1.0     0.266006      0.056016      0.481920     Biso  1.000000 H 
   H5         1.0     0.056014      0.733987      0.518061     Biso  1.000000 H 
   H6         1.0     0.233994      0.556016      0.481920     Biso  1.000000 H 
   H7         1.0     0.943986      0.266013      0.518061     Biso  1.000000 H 
   H8         1.0     0.766006      0.443984      0.481920     Biso  1.000000 H 
   O1         1.0     0.722266      0.777758      0.499997     Biso  1.000000 O 
   O2         1.0     0.277734      0.222242      0.499997     Biso  1.000000 O 
   O3         1.0     0.222266      0.722242      0.499997     Biso  1.000000 O 
   O4         1.0     0.777734      0.277758      0.499997     Biso  1.000000 O 

#======================================================================

# x. Structure R, Pressure 2 GPa, revPBE-vdW

data


_pd_phase_name                         'H O                                   ' 
_cell_length_a                         5.95637 
_cell_length_b                         4.77093 
_cell_length_c                         20.00000 
_cell_angle_alpha                      90 
_cell_angle_beta                       90 
_cell_angle_gamma                      90.00133 
_symmetry_space_group_name_H-M         'P 1' 
_symmetry_Int_Tables_number            1 

loop_ 
_symmetry_equiv_pos_as_xyz 
   'x, y, z' 

loop_ 
   _atom_site_label 
   _atom_site_occupancy 
   _atom_site_fract_x 
   _atom_site_fract_y 
   _atom_site_fract_z 
   _atom_site_adp_type 
   _atom_site_B_iso_or_equiv 
   _atom_site_type_symbol 
   H1         1.0     0.464354      0.229049      0.515424     Biso  1.000000 H 
   H2         1.0     0.259231      0.038345      0.533838     Biso  1.000000 H 
   H3         1.0     0.964390      0.229027      0.484583     Biso  1.000000 H 
   H4         1.0     0.759139      0.038458      0.466180     Biso  1.000000 H 
   H5         1.0     0.035586      0.729024      0.484567     Biso  1.000000 H 
   H6         1.0     0.240688      0.538263      0.466174     Biso  1.000000 H 
   H7         1.0     0.740763      0.538346      0.533797     Biso  1.000000 H 
   H8         1.0     0.535617      0.728996      0.515428     Biso  1.000000 H 
   O1         1.0     0.393949      0.044352      0.506561     Biso  1.000000 O 
   O2         1.0     0.893932      0.044342      0.493429     Biso  1.000000 O 
   O3         1.0     0.105996      0.544374      0.493465     Biso  1.000000 O 
   O4         1.0     0.606011      0.544337      0.506531     Biso  1.000000 O 
\end{verbatim}
}
\end{widetext}

\end{document}